\documentclass[onecolumn, amssymb, amsmath, superscriptaddress, eqsecnum, nofootinbib, notitlepage]{revtex4-1}
\usepackage{graphicx, braket}
\usepackage[colorlinks=true, citecolor=blue, linkcolor=., anchorcolor=.,filecolor=.,menucolor=.,runcolor=.,urlcolor=.,]{hyperref}
\usepackage[english]{babel}
\usepackage{verbatim}
\usepackage{tikz,bbm,bm}

\newcommand{\bigtimes}{\mathbin{\tikz [x=1.4ex,y=1.4ex,line width=.2ex] \draw (0,0) -- (1,1) (0,1) -- (1,0);}}%

\newcommand{\be}{\begin{equation}}
\newcommand{\ee}{\end{equation}}

\newcommand{\h}{\hat}

\newcommand{\ex}{\text{ex}}
\newcommand{\mcO}{\mathcal{O}}
\newcommand{\su}{\mathfrak{su}}
\newcommand{\N}{\mathfrak{N}}
\newcommand{\n}{\mathfrak{n}}
\newcommand{\Tr}{\text{Tr}}
\newcommand{\mcH}{\mathcal{H}}

\newcommand{\mbP}{\mathbb{P}}
\newcommand{\mbC}{\mathbb{C}}
\newcommand{\mbI}{\mathbb{I}}
\newcommand{\mbQ}{\mathbb{Q}}

\makeatletter
\newcommand*\bigcdot{\mathpalette\bigcdot@{.5}}
\newcommand*\bigcdot@[2]{\mathbin{\vcenter{\hbox{\scalebox{#2}{$\m@th#1\bullet$}}}}}
\makeatother

\begin{document}

\title{Statistical equilibrium of tetrahedra from maximum entropy principle}

\author{Goffredo Chirco} 
\email{goffredo.chirco@aei.mpg.de}
\affiliation{Max Planck Institute for Gravitational Physics (Albert Einstein Institute), Am M\"{u}hlenberg 1, 14476 Potsdam-Golm, Germany.}
\author{Isha Kotecha}
\email{isha.kotecha@aei.mpg.de}
\affiliation{Max Planck Institute for Gravitational Physics (Albert Einstein Institute), Am M\"{u}hlenberg 1, 14476 Potsdam-Golm, Germany.}
\affiliation{Institute for Physics, Humboldt-Universit\"{a}t zu Berlin, Newtonstra{\ss}e 15, 12489 Berlin, Germany.}
\author{Daniele Oriti}
\email{daniele.oriti@aei.mpg.de}
\affiliation{Max Planck Institute for Gravitational Physics (Albert Einstein Institute), Am M\"{u}hlenberg 1, 14476 Potsdam-Golm, Germany.}
\affiliation{Arnold-Sommerfeld-Center for Theoretical Physics, Ludwig-Maximilians-Universit\"at \\ Theresienstrasse 37, D-80333 M\"unchen, Germany.}

\begin{abstract}

Discrete formulations of (quantum) gravity in four spacetime dimensions build space out of tetrahedra. We investigate a statistical mechanical system of tetrahedra from a many-body point of view based on non-local, combinatorial gluing constraints that are modelled as multi-particle interactions. We focus on Gibbs equilibrium states, constructed using Jaynes' principle of constrained maximisation of entropy, which has been shown recently to play an important role in characterising equilibrium in background independent systems. We apply this principle first to classical systems of many tetrahedra using different examples of geometrically motivated constraints. 
Then for a system of quantum tetrahedra, we show that the quantum statistical partition function of a Gibbs state with respect to some constraint operator can be reinterpreted as a partition function for a quantum field theory of tetrahedra, taking the form of a group field theory. %Particularly, its classical contribution can be seen as defining a group field theory partition function.

%As an example of this principle in constrained systems, a Gibbs state with respect to the closure constraint for polygons is investigated. This state is found to be a generalisation of Souriau's Gibbs states to the case of a first class constraint. Applying this principle to a system of finitely many classical tetrahedra, we understand a Gibbs state with respect to area-matching gluing constraints as a statistical mixture of configurations of which those that admit a twisted geometry configuration are weighed exponentially more. In this sense such a state describes a system that is approximately twisted geometric. 

\end{abstract}

\maketitle

\tableofcontents

\newpage

\section{Introduction}

General relativity has taught us that gravity \emph{is} spacetime geometry, and its observational successes are testimony of the fruitfulness of this lesson, beyond its purely aesthetic appeal. Modern theoretical physics, however, has also provided hints that this continuum description of spacetime and gravity could be emergent, and that some kind of discrete substratum may replace it at the fundamental level \cite{Oriti:2018dsg}. Taking these lessons seriously in the search for quantum gravity has led to non-perturbative, discrete frameworks that aim at constructing quantum theories of geometry, and at showing the emergence of continuum spacetime and GR from discrete foundations. A crucial ingredient in these are geometric objects like polyhedra that can be understood as quantum excitations of geometry. Canonical quantisation of general relativity using Ashtekar variables has led to spin network states \cite{Ashtekar:2017yom, Ashtekar:2004eh, Bodendorfer:2016uat}, which admit an interpretation in terms of geometric polyhedra \cite{Bianchi:2010gc,Baratin:2010nn}. Cellular complexes of the same type are also the underpinning of covariant spin foam models \cite{Perez:2012wv,Perez:2012db}, which have polyhedra dual to spin networks forming their boundary states. In fact, simplicial discretisations have been considered often, originally by Regge \cite{Regge:1961px} with the aim of providing a coordinate-free description of classical spacetime, and are the fundamental mathematical structures of simplicial quantum gravity approaches, like quantum Regge calculus \cite{Hamber:2009mt} and (causal) dynamical triangulations \cite{Ambjorn:2012jv}. Finally, the group field theory framework \cite{Oriti:2006se,Oriti:2014uga,Krajewski:2012aw} treats polyhedra quite literally as the quanta of spacetime by defining for them a quantum field theory whose interactions represent their gluing and evolution processes; and in doing so, it provides a reformulation of both loop quantum gravity and spin foam models, and of simplicial quantum gravity approaches.\

A fully background independent (quantum) statistical mechanical framework \cite{Rovelli:1993ys, Connes:1994hv, Rovelli:2012nv, Montesinos:2000zi, Chirco:2013zwa, kotecha} could be the best way to provide a foundation, and subsequently to analyse such discrete quantum gravity approaches, concerning in particular the emergence of spacetime structures in a continuum approximation \cite{Oriti:2018dsg}, treating spacetime itself as a (peculiar) quantum many-body system made of tetrahedra \cite{Oriti:2017twl}. The definition of such framework poses however many challenges, starting from the identification of a good notion of equilibrium states. Let us say a few words on some of these challenges, and on previous work tackling them. \

Presently we are interested in defining Gibbs equilibrium states for a system of an arbitrary but finite number of tetrahedra, with respect to certain gluing constraints motivated from considerations in discrete quantum gravity. As is immediately evident, in such a system there does not exist any notion of a time variable, which begs the question: what notions of equilibrium can a system of many polyhedra admit? In a recent work \cite{Kotecha:2018gof}, a statistical mechanics for simplicial degrees of freedom is defined, using the tools provided by a group field theory many-body representation of the same. Therein, general construction schemes are discussed for defining Gibbs states in background independent settings, relevant for both classical and quantum sectors, and independent of any specific underlying framework in which the system is defined. It is suggested that in addition to the Kubo-Martin-Schwinger (KMS) condition \cite{robinson}, the principle of constrained maximisation of entropy as proposed by Jaynes \cite{Jaynes:1957zza, Jaynes:1957zz}, could be the crucial one in such settings \cite{Kotecha:2018gof, kotecha}, allowing for greater generality. The explicit examples provided there however make extensive use of the technical advantages offered by the group field theory formalism. For instance, as an elementary illustration of the utility of Jaynes' principle, an example of a group field theory Gibbs state with respect to a geometric volume operator was presented, and found to naturally support Bose-Einstein condensation to a low-spin phase \cite{Kotecha:2018gof}. In this paper, we tackle the issue of constructing equilibrium states for a system of many tetrahedra on the basis of this principle and in much greater generality, at both classical and quantum level, for general constraints (but giving concrete examples based on a number of geometrically motivated ones), and without relying on specific discrete gravity approaches (but we will see that the group field theory framework emerges naturally in the quantum setting).\

The paper is organised as follows. We begin with a discussion of the principle of maximum entropy a la Jaynes while emphasising its role in background independent systems in section \ref{maxent}. Focussing first on a system of classical tetrahedra, section \ref{tet} presents its mechanics and statistical mechanics. Disconnected tetrahedra are modelled as `particles' and its mechanical model is defined via generically non-local, combinatorial `interactions'. These are gluing constraints in general, as encountered in discrete gravity literature. As a first illustrative example of using the maximum entropy principle in the context of a constrained system related to tetrahedra, we study the case of closure constraint for a single open tetrahedron and construct a Gibbs state with respect to it in section \ref{entclosure}. We find that such a state encodes the constraint information partially in a statistical way. Moreover, it is found to be a generalisation of Souriau's Gibbs states to the case of a first class constraint. We define the corresponding statistical system of many closed tetrahedra in sections \ref{cm} and \ref{manytet}, and consider gluing constraints for the same which can be interpreted as a definition of their dynamics or \lq interactions\rq. The result of imposing this set of constraints exactly is a labelled triangulation. We use the twisted geometry interpretation of the same constraints to illustrate further the way discrete geometry is (or could be) encoded in the system and to suggest further developments. Section \ref{quant} discusses the analogous system of many quantum tetrahedra, first outlining its mechanics and statistical mechanics, and subsequently showing that the quantum statistical partition function of a Gibbs state of such a system can be recast in the form of a quantum field theory of tetrahedra, and we show its relation to the group field theory framework.

%-------------------------------------------------
%-------------------------------------------------
\section{Generalised Gibbs state from entropy maximisation} \label{maxent}

Equilibrium statistical states are known to play an important role in macroscopic systems with a large number of micro-constituents. Particularly Gibbs thermal states are ubiquitous in physics, being utilised across the spectrum of fields, ranging from phenomenological thermodynamics, condensed matter physics, optics, tensor networks and quantum information, to gravitational horizon thermodynamics, AdS/CFT and quantum gravity. They are the completely passive states, stationary under the dynamics of the system, and maximising the system's entropy for a given energy. They are a special case of the more general algebraic KMS states, which encode a complete notion of equilibrium for arbitrarily large systems. In fact, Gibbs states are the \emph{unique} KMS states for finite systems. 

Even close-to-equilibrium systems can be modelled via the notion of local equilibrium in which various subsystems of the whole are taken to be locally at equilibrium, while the global dynamics is not stationary. Collective variables, such as number and temperature densities, then vary smoothly across these different patches, while having constant (equilibrium) values within a given subsystem. This basic idea underlies several techniques for coarse-graining in general, and as such displays again the usefulness of statistical equilibrium descriptions.

Thus also in discrete quantum gravity, statistical equilibrium states will be of value in ongoing efforts to get an emergent spacetime, based on techniques from finite-temperature quantum and statistical field theory. Since equilibrium statistical mechanics provides a theoretical footing for thermodynamics, such a framework for (candidate) quantum gravity degrees of freedom will also facilitate identification of thermodynamic variables with geometric interpretations, rooted in the underlying fundamental theory, to then make contact with studies in spacetime thermodynamics.

\subsection{Generalised Gibbs equilibrium}
As proposed by Jaynes in two seminal papers \cite{Jaynes:1957zza, Jaynes:1957zz}, given our limited knowledge of a system with many underlying degrees of freedom in terms of a set of observable averages $\{\langle \mcO_a \rangle = U_a\}$, the least biased statistical distribution (in the sense of not assuming more information about the system than what we actually have) over the microscopic state space of the system is obtained by maximising the information entropy of the said system. By doing this, we are using \emph{only} the amount of information we have access to, not less or more. The resulting distribution is of the Gibbs form, and faithfully encodes our knowledge (and lack thereof) of the microscopics of the system, thus it is the best one can do in order to infer other observable equilibrium properties of the same system. \

Consider a finite set $\{\mcO_a\}_{a=1,2,...}$ of smooth functions $\mcO_a : \Gamma_\ex \to \mathbb{R}$, on a finite-dimensional extended\footnote{By extended we mean that: 1) it is the unconstrained phase space with respect to any constraints under consideration; 2) the system is not equipped with any external time or clock variable, and even if such a variable exists then at this fully parametrized level it is one of the dynamical variables included in the definition of this phase space.} symplectic phase space $\Gamma_\ex$, with Liouville measure $d\lambda$. A statistical (density) state on a phase space is a real-valued, positive and normalised function on it with respect to the measure. Let $\rho$ be a statistical state on $\Gamma_\ex$, such that the statistical averages of $\mcO_a$ in $\rho$ are fixed,
\be \label{const1}
\langle \mcO_a \rangle_\rho \equiv \int_{\Gamma_\ex} d\lambda \; \mcO_a \,\rho \;\;=\;\; U_a \;,
\ee
assuming that the integrals are convergent so that $U_a$ are well-defined. The state $\rho$ is normalised by definition, so that
\be \label{const2}
 \langle 1 \rangle_\rho = 1\;,
\ee
and, its Shannon entropy is
\be \label{ent}
S[\rho] = -\langle \ln \rho \rangle_\rho \;.
\ee
Consider maximisation of $S[\rho]$ under the given set of constraints \eqref{const1} and \eqref{const2} \cite{Jaynes:1957zza}. Using the Lagrange multipliers technique, this amounts to finding a stationary solution for the following auxiliary functional
\be \label{auxfn}
L[\rho,\beta_a,\kappa] = S[\rho] - \sum_{a} \beta_a (\langle \mcO_a \rangle_{\rho} - U_a) - \kappa(\langle 1 \rangle_\rho - 1)
\ee 
where $\beta_a, \kappa \in \mathbb{R}$ are Lagrange multipliers. Then, requiring stationarity\footnote{Notice that requiring stationarity of $L$ with variations in Lagrange multipliers implies fulfilment of the constraints \eqref{const1} and \eqref{const2}. These two `equations of motion' of $L$ along with the one determining $\rho$ (coming from stationarity of $L$ with respect to $\rho$) provide a complete description of the system at hand.} of $L$ with respect to variations in $\rho$ gives a generalised Gibbs state
\be \label{genstate1}
 \rho_{\{\beta_a\}} = \frac{1}{Z_{\{\beta_a\}}} e^{-\sum_a \beta_a \mcO_a} \ee 
 with partition function
 \be Z_{\{\beta_a\}} \equiv \int_{\Gamma_\ex} d\lambda \; e^{-\sum_a \beta_a \mcO_a}  = e^{1+\kappa} \ee 
where as is usual, normalisation multiplier $\kappa$ is a function of the rest. The parameters $\{\beta_a\}$ are such that the partition function integral converges.\

Analogous arguments hold for finite quantum systems and the above scheme can be implemented directly \cite{Jaynes:1957zz}, as long as the operators under consideration are such that the relevant traces are well-defined on a kinematic (unconstrained) Hilbert space. Statistical states are density operators, i.e. self-adjoint, positive and trace-class operators, on the Hilbert space. Statistical averages for self-adjoint observables  $\h{\mcO}_a$ are now,
 \be \langle \h{\mcO_a} \rangle_\rho \equiv \Tr(\h{\rho} \, \h{\mcO}_a) = U_a \;. \ee
Following the constrained optimisation problem presented above gives a resultant Gibbs density operator,
 \be  \h{\rho}_{\{\beta_a\}} = \frac{1}{Z_{\{\beta_a\}}} e^{-\sum_a \beta_a \h{\mcO}_a} \ee
where the state is well-defined as long as the trace for the partition function converges. The consequence of non-commuting observables on the operational understanding of Jaynes' prescription as starting from a known thermodynamic state given by set of observable averages requires further investigation, particularly in covariant systems. See \cite{Jaynes:1957zz} for some discussions. \

The significance of the maximum entropy principle is its applicability to a wide variety of situations. As long as the mathematical description of a given system (in terms of a state space and an observable algebra) is well-defined, and we have access to certain observables $\{\mcO_a\}$ with which we can define a macrostate $\{U_a\}$ of the system, the maximum entropy principle can be applied to characterise a notion of statistical equilibrium. Already, the notion of equilibrium is implicit in the existence of the constraints \eqref{const1} which basically say that the system has certain properties that act as good observables to label the state of the system with, because their expectation 
%observed\footnote{In the common context of standard thermodynamic systems, observed values are the usual empirically measurable properties of the system. In the less common context of say discrete quantum gravity systems, we can understand observed values as expectation values of operators corresponding to those properties of the system which are in principle empirically measurable at relevant (macroscopic) scales.} 
values remain constant. This feature of remaining constant, which is taken as a starting point of this procedure, could have pointed us already to the fact that there can exist a certain equilibrium description of the system in terms of these variables. As emphasised above, what Jaynes' procedure does is to allow us to find this description in a way that is least biased.

\subsection{Vector-valued temperature}

The generalised Gibbs state in \eqref{genstate1} defines a unique equilibrium distribution labelled by a set of temperatures $\{\beta_a\}$. In fact, we can encode $\{\beta_a\}$ in a multi-component, real vector-valued inverse temperature $ \beta \equiv \{\beta_a\}$ and rewrite this state as 
\be \label{genstate}
 \rho_{\beta} = \frac{1}{Z_{\beta}} e^{- \beta \cdot \mcO} \ee
with the vector-valued function $\mcO=\{ \mcO_a\}$ accordingly defined.

In general, whenever $\beta \cdot \mcO$ is convex, $\{\beta_a\}$\footnote{Each individual `temperature' $\beta_a$ defines the periodicity in the flow of $\mcO_a$.} can be understood as generalised (inverse) temperatures, determined by the constraints \eqref{const1}, via the equations
\be
-\frac{\partial \ln {Z_\beta}}{\partial \beta_a} = U_a
\ee
for each $a$. Other standard equilibrium thermodynamic relations follow, at least formally. In particular, the partition function $Z_{\{\beta_a\}}$, or equivalently the thermodynamic free energy potential 
\be \Phi_{\beta} := - \ln Z_{\beta} %\textcolor{blue}{ =-\sum_a \ln Z(\beta_a)= \sum_a \Phi_a}
\ee 
encodes complete thermodynamic information about the system. Quantities $U_a = \langle \mcO_a \rangle$ play the role of generalised energies, and  
 \be S = { \sum_a  \beta_a U_a  -  \Phi_{\beta}  }\ee is the entropy of state \eqref{genstate} such that,
 \be dS = \sum_a \beta_a (dU_a - \langle d\mcO_a \rangle)= \sum_a \beta_a (dU_a +dW_a)= \sum_a \beta_a \,dQ_a\;, \ee
where $ dW_a= -\langle d\mcO_a \rangle = \frac{1}{\beta_a} \int_{\Gamma_\ex} d\lambda \; \frac{\delta}{\delta \mcO_a}(\ln {Z_\beta})\,d\mcO_a$ correspond to work associated with the generalised energy changes, while the $dQ_a$ define the generalised heat \cite{Jaynes:1957zza, Jaynes:1957zz} variation of the system. 

%\textcolor{red}{To ensure physical equilibrium wrt dynamical constraint: $\{C,\mcO_a\} = 0 \Rightarrow \{C,e^{-\sum\beta_a \mcO_a}\}=0$. This way, the generic observables $\mcO$ defining the state are in fact gauge-invariant/conserved ones, as considered by Montesinos, Rovelli 2001.} 

\subsection{Modular flow, stationarity and global equilibrium}

Given a generalised Gibbs state as a result of the maximum entropy principle, one can (if one wants) extract a one-parameter modular flow, generated by $-\ln \rho$, in the sense that $\rho \, \omega(X_\beta)= \mathrm d \rho$, where $X_\beta$ is the modular vector field induced by the function $\beta \cdot \mcO$ (for the particular case above), which plays the role of a generalised modular Hamiltonian, while $\omega$ is the symplectic form on $\Gamma_{\ex}$. By construction, the Gibbs state will be stationary with respect to this flow. In this sense, any such $\rho$ always satisfies stationarity, which is the more traditional characterisation of equilibrium, with respect to its own modular flow\footnote{In fact, any faithful algebraic state over a von Neumann algebra defines a 1-parameter Tomita flow with respect to which the state satisfies the KMS condition. The deep significance of this fact for background independent systems was realised in \cite{Rovelli:1993ys, Connes:1994hv}.}. For cases when the Gibbs state is defined by generators of certain transformations, then the modular flow parameter is a rescaling of the parameter of the said transformations by a factor of $1/\beta$. In more general cases when $\mcO$ is not a generator of some transformation a priori, then the interpretation of its modular flow parameter would depend on the specific case at hand. 

Notice that in the special case when the set of constraints defining \eqref{genstate1} is completely independent, so that the set of the associated flows satisfy $[X_{a}, X_{a'}]= 0 \;\; \forall a, a'$, the full equilibrium state in addition to being stationary with respect to its modular flow, is also at equilibrium with respect to each $X_a$ separately. \ 

%can be understood as a product of equilibria and consistently written in a factorised form
%\be
%\rho_{\{\beta_a\}} = \prod_a \rho(\beta_a)= \prod_a \frac{1}{Z(\beta_a)} e^{- \beta_a \mcO_a}\ee
%where $Z_{\{\beta_a \}}= \prod_a Z(\beta_a)$. In this case, stationarity with respect to the evolution induced by $\rho_{\{\beta_a\}}$ is realised only in terms of stationarity with respect to the set of individual modular flows. 

A global notion of equilibrium characterised by a single temperature can be prescribed by coupling the individual flows \cite{Chirco:2013zwa}. This is a natural consequence of defining a Gibbs state according to Jaynes' procedure under the constraint $\langle h \rangle = 0$, where the generalised modular Hamiltonian $h = \sum_a \beta_a \mcO_a$ is in general a linear combination of the individual generators considered previously. Then, the resultant state $\rho_\beta = \frac{1}{Z_\beta}e^{-\beta h}$ is associated with a single $\mathbb{R}$-valued temperature $\beta$ which is related to the individual flows by,
\be \beta X_h = \sum_a \beta_a X_a \;.\ee
This essentially fixes the modular parameter $\tau$ (associated with $X_h)$ in the space of the flow parameters $t_a$ of the different $\mcO_a$ as, $ \tau = \sum_{a} \frac{\beta}{\beta_a} t_a $. 
Thus, to the state $\rho_{{\beta}}$ we can associate a one-parameter flow in $\Gamma_{\ex}$ given by a linear combination of the individual flows of $\mcO_a$. Extraction of a single \emph{physical} temperature would require additional inputs in terms of the physical interpretations of $\mcO_a$ and whether any one of them admits an understanding of energy. This will be investigated elsewhere.

%\textcolor{red}{$\beta_a$ understood better as coupling constants in the model-defining `modular hamiltonian' given by $h \equiv \sum_a \beta_a \mcO_a$. given a state \eqref{genstate}, then it corresponds to a single `temperature' which is the periodicity in the modular flow of $h$} 

\subsection{Remarks}

It was emphasised by Jaynes that this information-theoretic manner of defining equilibrium statistical mechanics (and from it, thermodynamics) is to elevate the status of entropy as being more fundamental than even energy. This perspective can be crucially appropriate in settings where energy (and time) are ill-defined or not defined at all. Moreover, this procedure is valid for both classical and quantum settings, and does not technically require any symmetries of the system to be defined a priori, unlike in the more traditional characterisation using the KMS condition. The maximum entropy principle could thus be particularly useful in background independent settings, including both covariant systems on spacetime, and more radical non-spatiotemporal systems like those in discrete quantum gravity (regardless of the specific framework) \cite{Kotecha:2018gof, kotecha}. \

Observables $\mcO$ which characterise a given equilibrium state, in principle, only need to be mathematically well-defined in the given description of the system. Valid choices include a Hamiltonian for time translations in non-relativistic systems \cite{Landau:1980mil}; a clock Hamiltonian for evolution in a reference matter variable in deparametrized systems \cite{Kotecha:2018gof}; geometric observables such as 3-volume \cite{Jaynes:1957zza, Kotecha:2018gof}; and gauge-invariant quantities in presymplectic systems \cite{Montesinos:2000zi}. Even generators of kinematic symmetries such as rotations, or more generally, of 1-parameter subgroups of Lie group actions \cite{souriau, e18100370} can be used. The key point is that: ({\emph{i}) these observables need not necessarily encode a physical model of the system and can correspond purely to structural properties such as a kinematic symmetry or a geometric aspect (i.e. they can be physical or structural); and ({\emph{ii}) they need not necessarily be generators of symmetries or physical evolution (including time translation), and can correspond to those properties that are not naturally associated to any transformations of the system (i.e. they can be dynamical or thermodynamical) \cite{Kotecha:2018gof}. Another feature that can be asked of observables $\mcO$ is that they be gauge-invariant, when gauge symmetries are present. This then ensures that the Gibbs state is defined on the reduced, gauge-invariant state space and in this sense is a physical statistical state of the system. Having clarified these general points, in this work we are mostly interested in gluing constraints in a system of many tetrahedra producing connected configurations and interpreted as a time-independent notion of \lq dynamics\rq, adapted to a discrete quantum gravity setting. 

%-------------------------------------------------

\section{Many classical tetrahedra}\label{tet}

\subsection{Statistical fluctuations in closure} \label{entclosure}

The Jaynes' characterisation of equilibrium allows for a natural group-theoretic generalisation of thermodynamics, whenever the constraint is associated to some (dynamical) symmetry of the system. In this case, the momentum map associated to the Hamiltonian action of the symmetry group on the covariant phase space of the system plays the role of a generalised energy function, comprising the full set of conserved quantities. Moreover, its convexity properties allow for a generalisation of the standard equilibrium thermodynamics \cite{souriau}. \ %Notice also that, in a relativistic framework, at least for first class constrained systems, dynamical constraints can be expressed in terms of the vanishing of the momentum map defined with respect to the reparametrization symmetry of the system, on the extended phase space \cite{}. ...}

This approach is useful also in our simplicial geometric context. We want to use generalised Gibbs states to define along these lines a statistical characterisation of the tetrahedral geometry in terms of its closure, starting from the extended phase space of a single open tetrahedron. The closure constraint is what allows to interpret geometrically a set of 3d vectors as the normal vectors to the faces of a polyhedron, and thus to fully capture its intrinsic geometry in terms of them. We will base on this our subsequent treatment of a system of many closed tetrahedra (or polyhedra in general).

%starting from the Kapovich-Millson moduli space description of classical polyhedron geometry \cite{kapovich1996}, and then base on this our treatment of the more general system of many polyhedra (particularly, tetrahedra).

%As a first step, we present a simple example of using the principle of maximum entropy in relation to the \emph{closure} constraint in the phase space $\mathcal{S}_d$ of classical geometries of a closed $d$-faced polyhedron with fixed face areas, up to rotations in 3d Euclidean space. 

Consider the symplectic space,
\begin{align}
\Gamma_{\{A_I\}} &= \{(X_I) \in \su(2)^{*4} \cong \mathbb{R}^{3\times 4} \;|\;  ||X_I|| = A_I\} \nonumber \\ &\cong S_{A_1}^2 \times ... \times S^2_{A_4} 
\end{align}   where each $S^2_{A_I}$ is a 2-sphere with radius $A_I$, and $I=1,2,3,4$. If the four vectors $X_I$ are constrained to sum to zero, the surfaces associated to them (as orthogonal to each of them) close, giving a $4$-polyhedron in $\mathbb{R}^3$ with faces of areas $\{A_I\}$ (see Figure \ref{convex}).\footnote{Analogous arguments hold for the case of an open $d$-polyhedron and its associated closure condition.} In this example, we consider $\Gamma_{\{A_I\}}$ as the extended phase space of interest, and denote $\Gamma_\ex \equiv \Gamma_{\{A_I\}}$. \

Consider then the diagonal action of the $SU(2)$ Lie group (rotations) on $\Gamma_\ex$. To this action we can associate a momentum map $J : \Gamma_\ex \to \su(2)^*$ defined by,
\be \label{mom}
J = \sum_{I=1}^4 X_I 
\ee
where $||X_I|| = A_I$. 
The symplectic reduction of $\Gamma_\ex$ with respect to the zero level set $J=0$, imposes closure of the four faces, resulting in the Kapovich-Millson phase space \cite{kapovich1996} $\mathcal{S}_4 = \Gamma_\ex//SU(2) = J^{-1}(0)/SU(2)$ of a closed tetrahedron with given face areas. Space $\Sigma \equiv J^{-1}(0)$ is the constrained submanifold. \ 

We are interested in defining an equilibrium state on $\Gamma_\ex$ by imposing the closure constraint (only) on average, along the lines described in \ref{maxent}. From a statistical perspective, we can interpret the exact, or `strong', fulfilment of closure as defining a microcanonical statistical state on $\Gamma_\ex$ with respect to this constraint, and therefore a generalised Gibbs state as encoding a `weak' fulfilment of the same constraint. These two states on the extended state space are formally related by a Laplace transform.   \

A Gibbs state with respect to closure for an open tetrahedron is defined by maximising the entropy functional \eqref{ent} under normalisation \eqref{const2} and the following three constraints,
\be \label{jconst}
\langle J_i \rangle_\rho \equiv \int_{\Gamma_\ex} d\lambda \; \rho \; J_i = 0 \hspace{0.5cm} (i=1,2,3)
\ee
where $\rho$ is a statistical state defined on $\Gamma_\ex$, and $J_i$ are components of $J$ in a basis of $\su(2)^*$. Notice that $J_i$ are smooth, real-valued\footnote{Real-valued because the algebra $\su(2)$ under consideration is a vector space over the reals.} scalar functions on $\Gamma_\ex$. These are the functions of interest which take on the role of quantities $\mathcal{O}_a$ used in \eqref{const1}. We stress again that equation \eqref{jconst}, for each $i$, is a weaker condition than imposing closure exactly by $J_i = 0$. Optimising $L$ of equation \eqref{auxfn} then gives a Gibbs state on $\Gamma_\ex$ of the form,
\be \label{closuregibbs}
\rho_{\beta} = \frac{1}{Z{(\beta)}} e^{-\beta \cdot J} 
\ee
where now the Lagrange multiplier $\beta$ is a vector in the algebra $\su(2)$, with components $\beta_i$, and $\beta \cdot J= \sum_{i} \beta_i J_i $ denotes its inner product with $J \in \su(2)^*$.
%$\{\beta_a  \in \mathbb{R}\}$ are Lagrange multipliers for the constraints \eqref{jconst}, and we have used the notation
%\be \label{innerprod} \beta \bigcdot J \equiv \sum_{a=1}^3 \beta_a J_a \;. \ee 
The equilibrium partition function is given by,
\be
Z(\beta) = \int_{\Gamma_\ex} d\lambda \; e^{-{\beta \cdot J}}
\ee
where $\beta$ is such that the integral converges. \

%Notice that $\beta \bigcdot J$ is a smooth, real-valued function on $\Gamma_\ex$, and the sum \eqref{innerprod} can be understood as an inner product between the dual algebra element $J(.) \in \su(2)^*$ and a vector $\beta \in \su(2) \cong \mathbb{R}^3$.
Now, recall $J$ being the momentum map corresponding to the diagonal action of $SU(2)$ on $\Gamma_\ex$. The corresponding co-momentum map $\beta \cdot J\equiv J(\beta)$ then plays the role of the modular Hamiltonian of the system on $\Gamma_\ex$. Therefore state $\rho_{\beta}$, constructed using the maximum entropy principle is in fact an example of a generalisation of the Gibbs states defined by Souriau \cite{souriau, e18100370}, to the case of Lie group actions associated to gauge symmetries generated by first class constraints. In this case, the vanishing of the associated momentum map is directly related to fulfilling the closure constraint. A detailed analysis of the single tetrahedron thermodynamics is given in \cite{chircoj}. \

The state $\rho_\beta$ encodes equilibrium with respect to translations along the integral curves of the vector field $\xi_\beta$ on $\Gamma_\ex$, defined by the equation $\omega(\xi_\beta) = -\mathrm d J(\beta)$, where $\omega$ is the symplectic 2-form on $\Gamma_\ex$. It is the fundamental vector field corresponding to vector $\beta \in \su(2)$. In other words, $\rho_\beta$ encodes equilibrium with respect to the one-parameter flow characterised by $\beta$, which is a generalised vector-valued temperature. This is analogous to the well-known case of accelerated trajectories on Minkowski spacetime, where thermal equilibrium can be established along Rindler orbits defined by the boost isometry, where $\beta$ defines the Unruh (inverse) temperature. Another example, in quantum gravity, is that of momentum Gibbs states constructed in group field theory \cite{Kotecha:2018gof}. 

%--------------------------------------------------------

\subsection{Classical mechanics and statistical mechanics}\label{cm}

As we have seen, we can encode the classical intrinsic geometry of a polyhedron by symplectically reducing, with respect to the closure condition, the space $\Gamma_\ex$, to get its Kapovich-Millson space \cite{kapovich1996},
\begin{align}
\mathcal{S}_d&= \{(X_I) \in \su(2)^{*d} \;|\; \sum_I X_I=0, ||X_I|| = A_I\} \nonumber \; .\end{align}
In general, the space $\mathcal{S}_d$} is a $(2d-6)$-dimensional symplectic manifold (Figure \ref{convex}). One could lift the restriction of fixed face areas, thereby adding $d$ degrees of freedom, to get the $(3d-6)$-dimensional space of closed polyhedra modulo rotations. For $d=4$, this is the 6-dim space of a tetrahedron \cite{Barbieri:1997ks, Baez:1999tk}, considered often in discrete quantum gravity contexts. This space corresponds to the possible values of the 6 edge lengths of a tetrahedron, or to the 6 areas of its four faces and two independent areas of parallelograms identified by midpoints of pairs of opposite edges. This space is not symplectic in general, and to get a symplectic manifold from it, one can either remove the $d$ area degrees of freedom to get $\mathcal{S}_d$, or add $d$ number of $U(1)$ degrees of freedom (angle conjugates to the areas), to get the spinor description of the so-called framed polyhedra \cite{Livine:2013tsa}. 
Along the lines showed in \ref{entclosure}, we can easily extend the statistical description to the case of the framed polyhedron system. However, we are presently more interested in extending the statistical description to a collection of many closed polyhedra. \

Let us then consider the space of closed polyhedra with a fixed orientation and extend the phase space description so to encompass the extrinsic geometric degrees of freedom, which we expect to play a role in the description of the coupling leading to a collective model. 

The face normal vectors can be seen as elements of the dual algebra $\su(2)^* \cong \mathbb{R}^3$, which is a Poisson manifold with its Kirillov-Kostant Poisson structure \cite{Baez:1999tk}. We add conjugate variables to these $\su(2)^*$ degrees of freedom (thereby, doubling the dimension) and consider the phase space, $T^*(SU(2)^d/SU(2))$, where the quotient by $SU(2)$ encodes the imposition of the closure.\footnote{In the context of gravity, other choices for the Lie group are $Spin(4)$ and $SL(2,\mathbb{C})$, which could be dealt with in our framework in an entirely analogous manner.} \

We further restrict to the case of tetrahedra $(d=4)$. Then the single `particle' classical phase space under consideration is,
\be \label{gammatet}
\Gamma = T^*(SU(2)^4/SU(2)) \;.
\ee 
The extended phase space of an $N$-particle classical system is given by the direct product space,
\be \label{gammaN}
\Gamma_N = \Gamma^{\times N} \;.
\ee
On this space mechanical models for a system of many tetrahedra can be defined via constraints among such tetrahedra. Typical examples would be non-local, combinatorial gluing constraints%\footnote{Here, by constraints we only mean certain relations between the underlying particle degrees of freedom, without necessarily relating to gauge symmetries.}
, possibly scaled by an amplitude weight. From the point of view of many-body physics, we expect these gluing constraints to be in fact modelled as generic multi-particle interactions, defined in terms of tetrahedron intrinsic and extrinsic geometric degrees of freedom. Different choices of these interactions identify different models of the system. %Such combinatorial `Hamiltonians' constraints define the `dynamics', or chosen constrained collective configurations, of a system of many tetrahedra. 

\begin{figure}[t]
\includegraphics[width=3.5 in]{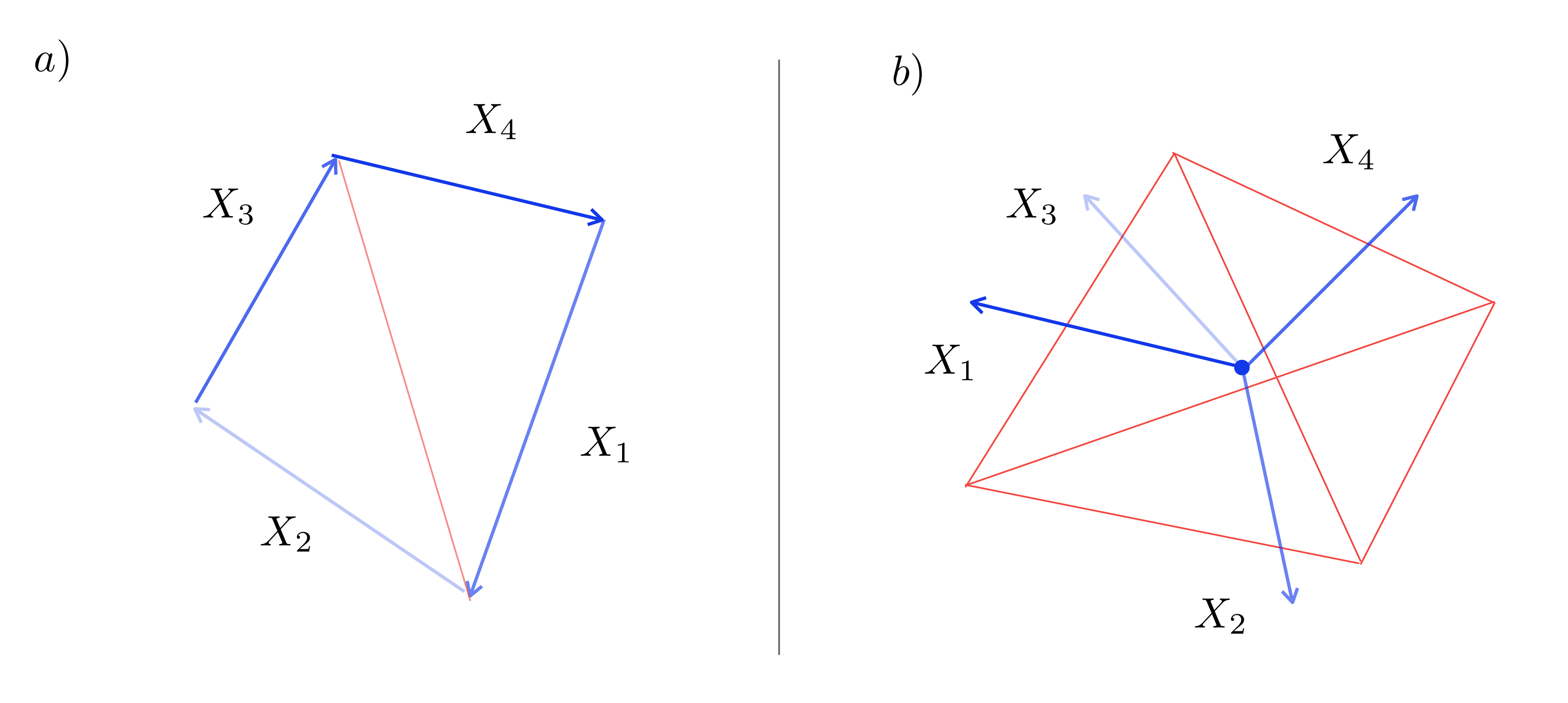}
\caption{(a) A convex polygon with side vectors $X_I$. The space of possible polygons in $\mathbb{R}^3$ up to rotations is a $(2d-6)$-dimensional phase space. For non-coplanar normals, the same data define also a unique polyhedron by Minkowski's theorem. (b) For $d=4$ we get a geometric tetrahedron. }\label{convex}
\end{figure}

Thus, the minimal interaction or, better, the key ingredient of such interactions of any many-tetrahedra model are the constraints which glue two faces of any two different tetrahedra. By gluing we mean here the requirement that the areas of these faces match and that their face normals align (with opposite orientation). We will detail below how these conditions are implemented. More stringent conditions, imposing stronger matching of geometric data, as well as more relaxed ones, can also be considered, as will be discussed below. What constitutes as gluing is thus a model-building choice, and so is the choice of which combinatorial pattern of gluings among a given number of tetrahedra is enforced. So, once the system knows how to glue two faces, then the remaining content of a model dictates how the tetrahedra interact non-locally, face-wise, to make simplicial complexes. Again, we will show some such choices below, when illustrating examples of our general framework. \

 %(see \cite{Baytas:2018wjd} for a nice overview). \footnote{In a system of finitely many classical tetrahedra, area-matching of shared faces, (anti-)aligning orientations of the two tetrahedra sharing these faces and shape-matching allow for interpretations in terms of discrete generalised geometries, namely twisted geometries, vector geometries and Regge geometries respectively (see \cite{Baytas:2018wjd} for a nice overview and references). Twisted geometries are also discussed in section \ref{gibbstwisted} in this paper.} 

 %From the point of view of a multi-particle interacting system, such a situation\footnote{This corresponds to loop diagrams in the dual labelled graph or spin network.} corresponds to including self-interactions. \

 %The dimensionality of the resulting simplicial complex depends crucially on the combinatorics encoded in the constraints defining the model. In the context of discrete gravity \textcolor{orange}{(cite?)}, an important model-building choice is whether to allow disconnected pieces of simplicial complexes to be generated dynamically. Another is to impose additional simplicity \textcolor{orange}{(?)} constraints on a 3d triangulation to allow for interpreting it as a boundary state of a 4d simplicial complex \textcolor{orange}{(cite?)}. 

An outline of the ensuing statistical system is as follows. A mechanical model of a system of many classical tetrahedra thus consists of a state space \eqref{gammaN}, an algebra of smooth functions over it and a set of gluing constraints defining the constrained dynamics. Further, a statistical mechanical model is defined by a statistical state (a real-valued, positive and normalised function) on this same system.  And equilibrium configurations comprised of collections of geometric tetrahedra can be constructed, at least formally, by using Jaynes' principle in terms of a suitable set of gluing constraints for such mechanical models. \

To consider then a system of an arbitrary, variable number of tetrahedra in a statistical setting amounts to including grand-canonical type probability weights, $e^{\mu N}$.  Let $Z_N$ be the partition function, which encodes (by definition) all statistical and thermodynamical information about the state $\rho_N$ on $\Gamma_N$, $Z_N = \int_{\Gamma_N} d\lambda \; \rho_N$. Then a system with a variable (and arbitrary, possibly infinite) particle number is described by, $Z = \sum_{N \geq 0} e^{\mu N} Z_N$. 

%--------------------------------------------------------

\subsection{System of tetrahedra at equilibrium}\label{manytet}

We now detail the construction of classical Gibbs states for systems of many classical tetrahedra with some concrete examples. The key ingredient is a set of conditions, the \lq gluing conditions\rq, which are understood as the constraints which lead from a set of disconnected tetrahedra to an extended simplicial complex. The same gluing process can be encoded in terms of dual graphs, understood as the 1-skeleton of the cellular complex dual to the simplicial complex of interest. The geometry of the initial set of tetrahedra as well as of the resulting simplicial complex can be captured by the $T^*SU(2)$ data introduced above. We will perform our construction in terms of these data first. A more detailed, thus transparent, characterisation of the same (loose notion of) geometry can be obtained in terms of so-called twisted geometry decomposition, which we will connect with at a second stage, to suggest further research directions  based on our construction. \

%We first recall the twisted geometries parametrisation and clarify how it can be modelled as a certain class of combinatorial interactions between tetrahedra from the perspective of a many-body system. \

Let $\gamma$ denote an oriented, 4-valent closed graph with $L$ number of oriented links and $N$ number of nodes. Each link $\ell$ is dressed with $T^*SU(2) \cong SU(2) \times \su(2)^* \ni (g_\ell,X_\ell)$ data, with variables satisfying invariance under diagonal $SU(2)$ action at each node $n$. $\gamma$ is dual to a simplicial complex $\gamma^*$, with triangular faces $\ell$ and tetrahedra $n$. Geometric closure of each tetrahedron corresponds to $SU(2)$-invariance at the dual node. The source and target nodes (tetrahedra) sharing a directed link (face) $\ell$ are denoted by $s(\ell)$ and $t(\ell)$ respectively. A state $(g_\ell,X_\ell)$ on $\gamma$ is then an element  of $\Gamma_\gamma = T^*SU(2)^L//SU(2)^N$. Such configurations admit a loose notion of discrete geometry in terms of area vectors, normal to the surfaces dual to the links, and identifying a simplicial complex, as we have discussed above. The geometry so-defined is potentially pathological, in the sense that the resulting simplicial complex may not be fully specified in terms of metric data, i.e. its associated edge lengths, as a Regge geometry would be. For our purposes, though, this characterisation suffices to show how a statistical state can be constructed based on encoding gluing and possibly other constraints on the initially disconnected tetrahedra. We will discuss further the purely geometric aspects in the following.

\begin{figure}[t]
\includegraphics[width=3.5 in]{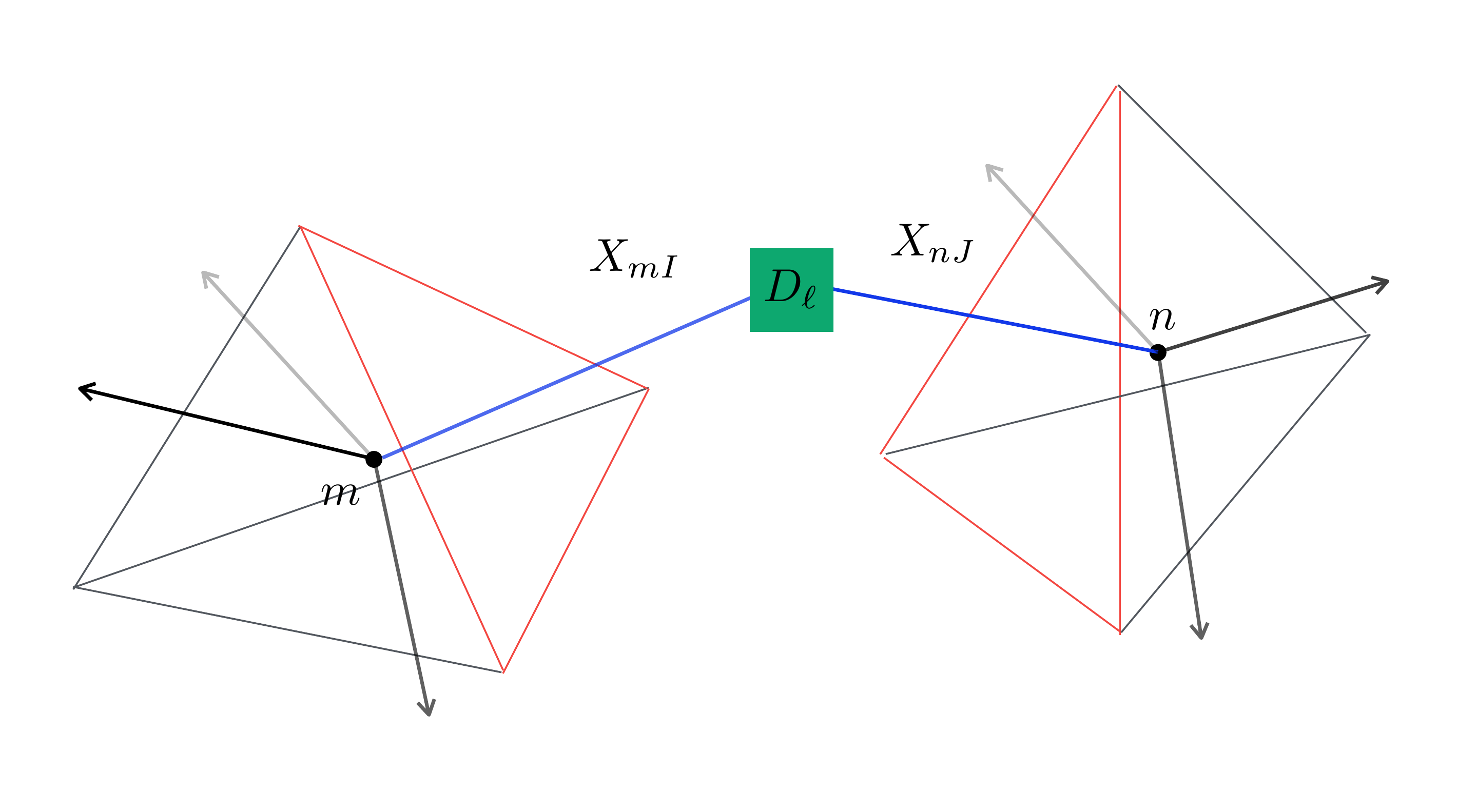}
\caption{Gluing via constraint $X_{(nJ)}+X_{(mI)} = 0$. }\label{gluings}
\end{figure}

To understand better the gluing process, and the resulting constraints, let us begin with a single closed, classical tetrahedron $n$ with state space $\Gamma$ of equation \eqref{gammatet}. As mentioned earlier, $\Gamma$ is the state space where 3d rotations have not been factorised out, which essentially means that each such tetrahedron is equipped with an arbitrary (orthonormal) reference frame determining its overall orientation in its $\mathbb{R}^3$ embedding. In the holonomy-flux representation, the four triangular faces $\ell |_n$, of tetrahedron $n$, are labelled by the four pairs $(g_\ell, X_\ell)$, with variables satisfying closure. 
In the dual picture, we have a single open graph node $n$ with four half-links $\ell |_n$ incident on it. Each half-link is bounded by two nodes, one of which is the central node $n$, common to all four $\ell$. Each $\ell$ is oriented outward (by choice of convention) from the common node $n$, which then is the source node for all four half-links. Then in the holonomy-flux parametrisation, each half-link $\ell$ is labelled by $(g_\ell,X_\ell)$. \

Let us denote the $I^{\text{th}}$ half-link belonging to an open node $n$ by $(nI)$, where $I=1,2,3,4$. Equivalently, $(nI)$ also denotes the $I^{\text{th}}$ face of tetrahedron $n$. 
Two tetrahedra $n$ and $m$ are said to be neighbours (Figure \ref{gluings}) if at least one pair of faces, $(nI)$ and $(mJ)$, are adjacent, that is the variables assigned to the two faces satisfy the following constraints,
\be \label{gluing} g_{(nI)}g_{(mJ)} = e\;\;\;,\;\;\; X_{(nI)}+X_{(mJ)}= 0 \;\;. \ee 

A given classical state associated to the connected graph $\gamma$ can then be understood as a result of imposing the constraints \eqref{gluing} on pairs of faces in a system of $N$ open nodes, or disconnected tetrahedra. That is, $\gamma$ is a result of imposing $L$ number each of $SU(2)$-valued and $\su(2)^*$-valued constraints, which we denote by $C$ and $D$ respectively. This in turn is a total $6L$ number of $\mathbb{R}$-valued (component) constraint functions $\{C_{\ell,a} \,, D_{\ell,a}\}$, for $\ell=1,2,...,L$ and $a=1,2,3$. For instance, creation of a full link $\ell = (nI,mJ)$ involves matching the fluxes, component-wise, by imposing the three constraints $D_{\ell,a} = X^a_{(nI)}+X^a_{(mJ)} = 0$, as well as restricting the conjugate parallel transports to satisfy $C_{\ell,a} = (g_{(nI)}g_{(mJ)})^a - e^a = 0$. Naturally the final combinatorics of $\gamma$ is determined by which half-links are glued pairwise, which is encoded in which specific pairs of such constraints are imposed on the initial data. \

As an example, consider a dipole graph, Figure \ref{dip}. This can be understood as imposing constraints on pairs of half-links of two open 4-valent nodes. Here $L=4$, thus we have at hand four constraints $D_\ell$ on flux variables,
\begin{align} \label{dipcon}
X_{(11)} + X_{(21)} &= 0 \;\;,	\quad	X_{(12)} + X_{(22)} = 0\;\;,	\\ \nonumber
X_{(13)} + X_{(23)} &= 0\;\;,	\quad	X_{(14)} + X_{(24)} = 0 \;\;.
\end{align}
This corresponds to a set of $3\times 4$ component constraint equations $D_{\ell,a}=0$. Similarly for holonomy variables. \ 

As another example, consider a 4-simplex graph made of five 4-valent nodes (Figure \ref{4sym}). The combinatorics is encoded in the choice of pairs of half links that are glued. Here $L=10$, corresponding to ten constraints $D_\ell$ on the flux variables, 
\begin{align} \label{4simp}
X_{(12)}+X_{(21)} &= 0 \;\;, & X_{(13)}+X_{(31)} &= 0 \;\;, & X_{(14)}+X_{(41)} &= 0 \;\;, & X_{(15)}+X_{(51)} &= 0 \;\;, & X_{(23)}+X_{(32)} &= 0 \;\;, \nonumber \\
X_{(24)}+X_{(42)} &= 0 \;\;, & X_{(25)}+X_{(52)} &= 0 \;\;, & X_{(34)}+X_{(43)} &= 0 \;\;, & X_{(35)}+X_{(53)} &= 0 \;\;, & X_{(45)}+X_{(54)} &= 0 \;\;.
\end{align}
As before, this corresponds to 30 component equations for the flux variables, and another 30 for holonomies. 

\begin{figure}
\centering
\begin{minipage}{.5\linewidth}
  \centering
  \includegraphics[width=3 in]{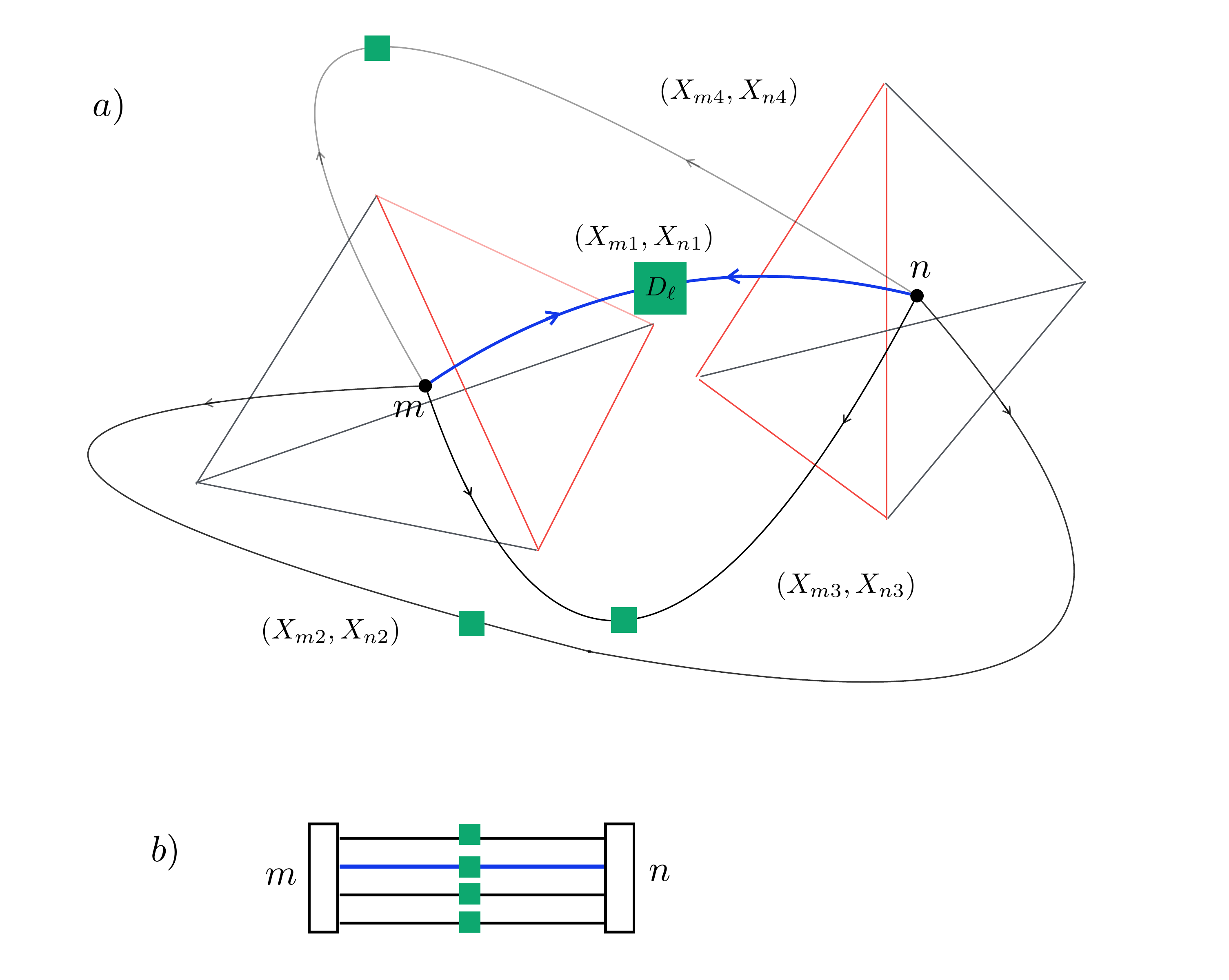}
\caption{(a) Dipole gluing in a system of two tetrahedra. \\ (b) Combinatorics of the dipole gluing.}\label{dip}
\end{minipage}%
\begin{minipage}{.5\linewidth}
  \centering
  \includegraphics[width=2.8 in]{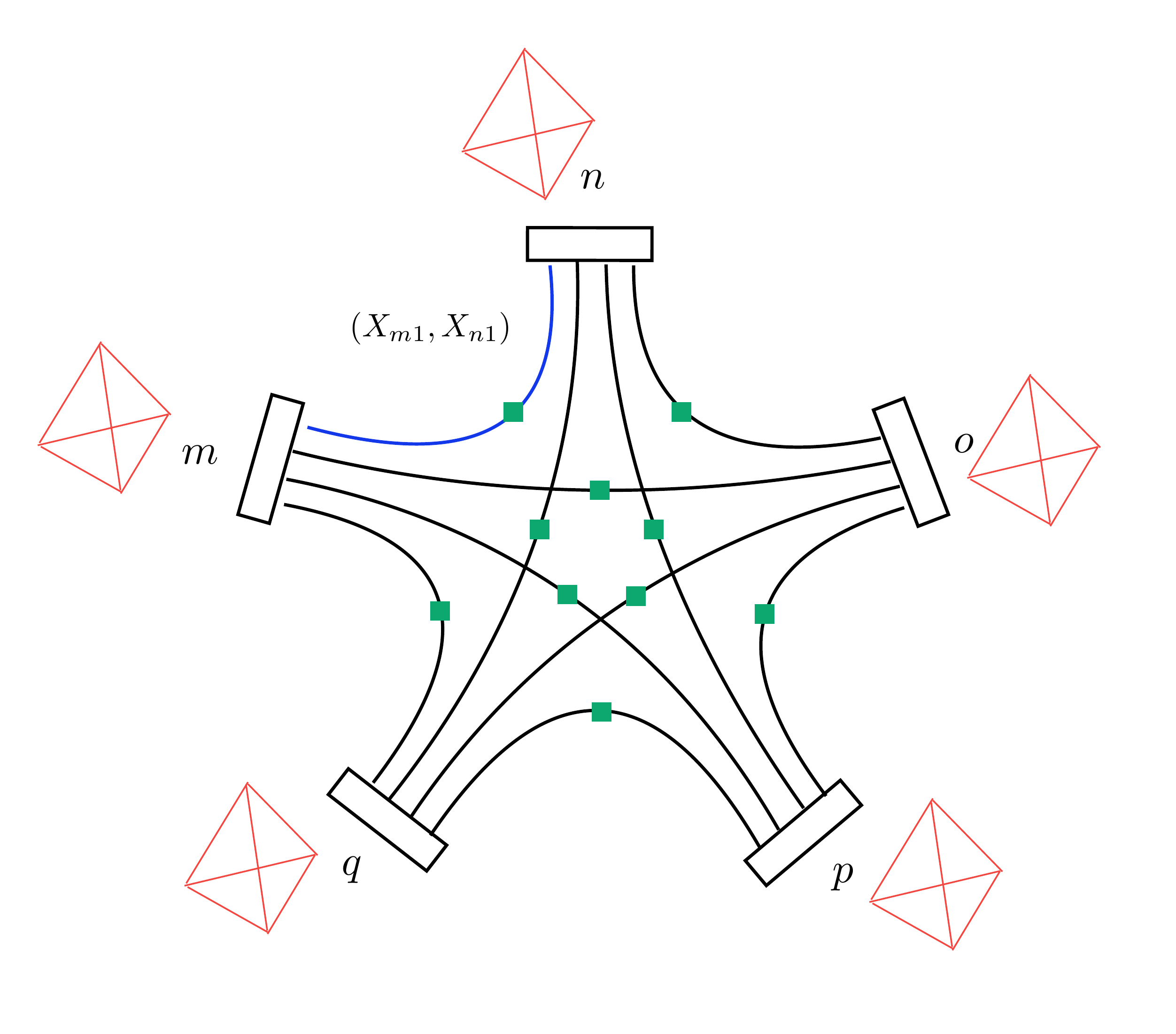}
  \caption{Resultant 4-simplex from combinatorial gluing between faces of five tetrahedra.}
  \label{4sym}
\end{minipage}
\end{figure}

When these constraints are satisfied exactly, that is $\{C_{\ell,a} = 0 \,, D_{\ell,a} = 0\}$ (for all $\ell, a$), then this system of $N$ tetrahedra admits a geometric interpretation based on the resultant simplicial complex. But as discussed in the previous section, there is a way of imposing these constraints only on average, that is $\{\langle C_{\ell,a} \rangle_\rho = 0 \,, \langle D_{\ell,a} \rangle_\rho = 0\}$. 

This statistical manner of weakly imposing the constraints results in a generalised Gibbs state, parametrised by $6L$ number of generalised temperatures, 
\be\label{rhocomp} \rho_{ \{\gamma, \alpha,\beta\}} \propto e^{-\sum_{\ell=1}^{L} \sum_{a=1}^3 (\alpha_{\ell,a} C_{\ell,a} + \beta_{\ell,a} D_{\ell,a})} \equiv e^{-{G}_\gamma ({\alpha},{\beta})}
 \ee
 where $\alpha,\beta \in \mathbb{R}^{3L}$ are vector-valued temperatures. Notice that the constraints $\{C_{\ell,a}, D_{\ell,a}\}$ are smooth functions on $\Gamma_N$, and $\rho_{ \{\gamma, \alpha,\beta\}}$ is thus a state defined for the $N$ particle system (assuming well-defined normalisation). In other words, the creation of a full link $\ell$ is associated with two $\mathbb{R}^3$-valued temperatures, $\alpha_\ell \equiv \{\alpha_{\ell,a}\}_{a=1,2,3}$ and $\beta_\ell \equiv \{\beta_{\ell,a} \}_{a=1,2,3}$. For instance, for the 4-valent dipole graph of Figure \ref{dip} with flux constraints \eqref{dipcon} and the corresponding holonomy ones, we have \be G_{\text{dip}}(\alpha,\beta) = \sum_{\ell=1}^4 \sum_{a=1}^3 \alpha_{\ell,a} C_{\ell,a} (g_{(1\ell)},g_{(2\ell)}) + \beta_{\ell,a} D_{\ell,a} (X_{(1\ell)},X_{(2\ell)})  \;. \ee \

We could further choose to assign a single temperature for all three components $a$. In such a case, one pair of $\mathbb{R}$-valued parameters, $\alpha_{\ell}$ and $\beta_{\ell}$, controls each link $\ell$, instead of three pairs. We would then have,
\be\label{rogam} \rho_{\{\gamma,\alpha,\beta\}} \propto e^{-\sum_{\ell=1}^L \alpha_\ell C_\ell + \beta_\ell D_\ell } \ee 
where $\alpha,\beta \in \mathbb{R}^L$, and $C_\ell = \sum_{a} C_{\ell,a}$. Making such different choices have non-trivial consequences. Notice that $\{C_{\ell,a}=0,D_{\ell,a}=0\}_{\gamma} \Rightarrow \{C_\ell=0,D_\ell=0\}_{\gamma}$ but the opposite is not true. Latter is thus a weaker condition than the former. States \eqref{rhocomp} and \eqref{rogam} correspond to these two set of conditions respectively, associated with constraints $\{\langle C_{\ell,a} \rangle = 0,\langle D_{\ell,a} \rangle = 0 \}$ and $\{\langle C_\ell \rangle = 0,\langle D_\ell \rangle = 0 \}$ in the entropy maximisation prescription.
%, as it depends on all individual link temperatures $\{\alpha_\ell , \beta_\ell\}$ which are the Lagrange multipliers for the individual constraints $\{\langle C_\ell \rangle = 0,\langle D_\ell \rangle = 0 \}$. 
If we were further to extract a single global temperature, say $\beta_\gamma$ (so that the state is of the form $\rho_{\beta_\gamma} \propto e^{-\beta_\gamma G_\gamma}$), then this would mean imposing the single condition $\langle G_\gamma \rangle = 0$, which in turn is weaker than the previous two sets (this being a sum of constraints). In these second and third weaker cases, the  corresponding microcanonical state cannot be understood as making the graph $\gamma$. In other words, to make $\gamma$ is to impose $\{C_{\ell,a}=0,D_{\ell,a}=0\}_{\gamma}$, and not either of the two weaker conditions.  \ 

%Notice that $\{C_{\ell,a}=0,D_{\ell,a}=0\}_{\gamma} \Rightarrow \{C_\ell=0,D_\ell=0\}_{\gamma} \Rightarrow G_\gamma = 0$, but the opposite sequence of implications does not hold. The latter conditions are thus progressively weaker than the former. States \eqref{rhocomp} and \eqref{rogam} correspond to the first and second set of conditions, associated with constraints $\{\langle C_{\ell,a} \rangle = 0,\langle D_{\ell,a} \rangle = 0 \}$ and $\{\langle C_\ell \rangle = 0,\langle D_\ell \rangle = 0 \}$ in the entropy maximisation prescription, respectively. If we were further to extract a single global temperature, say $\beta_\gamma$ (so that the state is of the form $\rho \propto e^{-\beta_\gamma G_\gamma}$), then this would mean imposing the single weaker condition $\langle G_\gamma \rangle = 0$. In these second and third weaker cases, the  corresponding microcanonical state cannot be understood as making the graph $\gamma$. In other words, to make $\gamma$ is to impose $\{C_{\ell,a}=0,D_{\ell,a}=0\}_{\gamma}$, and not either of the two weaker conditions.

%\textcolor{orange}{Why: Putting $\alpha_{\ell}=\beta_{\ell}$ sounds to strong to me. A priori each Lagrange multiplier is unique. How can we justify it?} \
 
A state such as \eqref{rhocomp} is a statistical mixture of configurations where the ones which are glued with the combinatorics of $\gamma$ (and thus admit a loose geometric interpretation) are weighted exponentially more than those which are not. In this sense it illustrates a statistical way to encode an approximate notion of discrete geometry. While states like \eqref{rogam} and $\rho_{\beta_\gamma}$ would encode a weaker notion of such statistical fluctuations, even if that, because in general even those configurations satisfying $\{C_\ell=0,D_\ell=0\}$ or $G_\gamma = 0$ exactly may not necessarily correspond to the graph $\gamma$. \\

The above can be generalised to include different interaction terms, each corresponding to a given pattern of gluings associated with a different graph $\gamma$. We first take into account all possible graphs $\{\gamma\}_N$ with a fixed number of nodes $N$. To each $\gamma$ in this set corresponds a gluing $G_\gamma(\alpha,\beta)$ as a function of several temperatures, which encode the fluctuations in the internal structure of the given graph. We can then think of a statistical mixture consisting of the different graphs represented by their respective combinatorics $G_\gamma$, each being weighted differently by coupling parameters $\lambda_\gamma$. Such a Gibbs distribution can be formally written as,
\be \label{rhon}
\rho_N = \frac{1}{Z_N(\lambda_\gamma,\alpha,\beta)} e^{-\sum_{\{\gamma\}_N} \frac{1}{\text{Aut}(\gamma)}\lambda_\gamma G_\gamma(\alpha,\beta)}
\ee
where Aut$(\gamma)$ factors out repetitions due to graph automorphisms. The choice of the set $\{\gamma\}_N$ is a model-building choice, analogous to choosing the different types of interaction potentials in standard many-body theory. \

Finally, if we allow for the number of tetrahedra $N$ to vary, we can write a general expression for the partition function of a system of many classical tetrahedra at equilibrium in a grand-canonical state,
 \be\label{grand} Z(\mu, \lambda_\gamma,\alpha,\beta) = \sum_{N} e^{\mu N}\,Z_N(\lambda_\gamma,\alpha,\beta)  \ee
where $\mu$ is the Lagrange parameter for $N$ and $Z _N$ is the canonical partition function for a finite number of tetrahedra, but including different graph contributions.

\
 
%\subsubsection*{Relation to twisted geometry}\label{gibbstwisted}

Now that we have presented the construction of a statistical state for many classical tetrahedra, that involves some set of gluing constraints, imposing a geometric interpretation, we can discuss briefly some model-building strategies that can be followed to construct more examples of interesting simplicial gravity models. Any such model building strategy should be based on a clear understanding of how simplicial geometry is encoded in the data we have used. \
 
A more precise parametrization of the holonomy-flux geometries can also be given in the language of twisted geometries \cite{Freidel:2010aq, Rovelli:2010km}. This relies on the fact that the link space $T^*SU(2)$ can be decomposed as $S^2 \times S^2 \times T^*S^1 \ni (\N_{s(\ell)},\N_{t(\ell)},A_\ell,\xi_\ell)$, modulo null orbits of the latter, and up to a $\mathbb{Z}_2$ symmetry. The variables are related by the following canonical transformations,  
\be
g = \n_s e^{\xi \tau_3} \n_t^{-1} \;\;\;,\;\;\; X = A \n_s \tau_3 \n_t^{-1}
\ee
where $\n_{s,t} \in SU(2)$ are those elements which in the adjoint representation $R$ rotate the vector $z \equiv (0,0,1)$ to give $\mathbb{R}^3$ vectors $\N_{s,t}$ respectively. That is $\N_{s,t} = R(\n_{s,t}).z$, or equivalently $\n_{s,t} \tau_3 \n_{s,t}^{-1} = \sum_{a=1}^3 \N^a_{s,t}\tau_a$ respectively for $s$ and $t$. Generators of $\su(2)$ are $\tau_a = -\frac{i}{2}\sigma_a$, where $\sigma_a$ are Pauli matrices. Vectors $\N_{s(\ell)}$ and $\N_{t(\ell)}$ are unit normals to the face $\ell$ as seen from two arbitrary, different orthonormal reference frames attached to $s(\ell)$ and $t(\ell)$ respectively. $A_\ell$ is the area of $\ell$, and $\xi_\ell$ is an angle which encodes (partial\footnote{The remaining two degrees of freedom of extrinsic curvature are encoded in the normals $\N_{s(\ell)}$ and $\N_{t(\ell)}$ \cite{Freidel:2010aq}. For instance, in the subclass of Regge geometries, $\xi_\ell$ is proportional to the modulus of the extrinsic curvature \cite{Rovelli:2010km}.}) extrinsic curvature information. So, a closed twisted geometry configuration supported on $\gamma$ is an element of $\bigtimes_{\ell} T^*S^1 \bigtimes_{n} \mathcal{S}_4$, where $\mathcal{S}_4$ is the Kapovich-Millson phase space of a tetrahedron given a set of face areas; each link is labelled with $(A_\ell, \xi_\ell)$, and each node with four area normals (in a given reference frame) that satisfy closure.\

A twisted geometry is in general discontinuous across the faces; so is the one described in terms of holonomy-flux variables, because both contain the same information. Face area $A_\ell$ of a shared triangle is the same as seen from tetrahedron $s(\ell)$ or $t(\ell)$ on either side; but the edge lengths when approaching from either side may different in general. That is, the shape of the triangle $\ell$, as seen from the two tetrahedra sharing it, is not constrained to match. If additional shape-matching conditions \cite{Dittrich:2008va} were satisfied, then we would instead have a proper Regge (metric) geometry on $\gamma^*$, which is a subclass of twisted geometries. These shape-matching conditions can be related to the so-called simplicity constraints, which are central in all model building strategies in the context of spin foam models, and whose effect is exactly to enforce geometricity (in the sense of metric and tetrad geometry) on discrete data of the holonomy-flux type, characterising (continuum and discrete) topological BF theories in any dimension. \

%In a classical system, as will be discussed in more detail later, these are implemented by equations \eqref{gluing}, and in a quantum setting by group (or flux) averaging of wavefunctions \cite{Oriti:2013aqa} as shown in equation \eqref{gluing2}. These faces could still be misaligned by a rotation around the face normal axis, and have different edge lengths in general. If by gluing we also mean rectifying these, then in addition to the area-matching condition \cite{Freidel:2010aq}, we must also impose additional constraints on the orientations of the two tetrahedra so that the faces are aligned properly edge-wise, and shape-matching conditions \cite{Dittrich:2008va} so that their edge lengths are the same.

%While in the twisted geometries parametrisation, each face $\ell$ of $n$ is labelled by $(A_\ell,\xi_\ell)$, along with two unit normals (in the given reference frame), $\N_{s(\ell)}$ and $\N_{t(\ell)}$, such that the four normals $\{\N_{s(\ell)}\}_{\ell |_n}$ satisfy closure. \

%While in the twisted geometries parametrisation, each $\ell$ is labelled by $(A_\ell,\xi_\ell)$; and its bounding nodes, $s(\ell)$ and $t(\ell)$, by unit normals $\N_{s(\ell)}$ and $\N_{t(\ell)}$ (in a given reference frame) respectively. Normals $\{\N_{s(\ell)}\}_{\ell |_n}$ satisfy closure. \

The gluing constraints in equation \eqref{gluing}, in twisted geometry variables, take the form of the following constraints
\begin{align} A_{(nI)} - A_{(mJ)} = 0 \;\;\;&,\;\;\; \xi_{(nI)} + \xi_{(mJ)} = 0 \;\;, \nonumber \\
  \N_{s(nI)} - \N_{t(mJ)} = 0 \;\;\;&,\;\;\;  \N_{t(nI)} - \N_{s(mJ)} = 0 \;. \label{gluingtwisted} \end{align}
The result is the same, of course, as in the holonomy-flux case: half-links $(nI)$ and $(mJ)$ which satisfy the above set of six component constraint functions (in either of the parametrisations) are thus glued\footnote{Gluing the two half-links is essentially superposing one over the other in terms of aligning their respective reference frames. This is evident from the constraints for the normal vectors $\N$ which superposes the target node of one half-link on the source node of the other, and vice-versa.} to form a single link $\ell \equiv (nI,mJ)$. Equivalently, the two faces of the initially disconnected tetrahedra are now adjacent. \

The more refined geometric data used in the twisted geometry language allow for a model-building strategy leading for example to statistical states in which only some of the gluing conditions are imposed strongly, while others are imposed only on average. In the same spirit of achieving greater geometrical significance of the statistical state that one ends up with, our construction scheme can be applied with additional constraints, beyond the gluing ones we illustrated above. For instance, starting with the space of twisted geometries on a given simplicial complex (dual to) $\gamma$, one could consider imposing (on average) also shape-matching constraints, or simplicity constraints, to encode an approximate notion of a Regge geometry using a Gibbs statistical state. This would be the statistical counterpart of the construction of spin foam models, i.e. discrete gravity path integrals in representation theoretic variables \cite{Perez:2012db, Baratin:2011hp, marco}, based on the formulation of gravity as a constrained BF theory. This is left to future work. \

%-------------------------------------------------

\section{Many quantum tetrahedra} \label{quant}

\subsection{Quantum mechanics and statistical mechanics} \label{tet2}
There are several ways of translating the classical construction presented in the previous sections at the quantum level, starting from a quantisation of the geometry of a single tetrahedron \cite{Barbieri:1997ks,Baez:1999tk}.

In a quantum setting in general, each closed polyhedron face $I$ is assigned an $SU(2)$ representation label $j_I$ with its associated representation space $\mathcal{H}_{j_I}$, and the polyhedron itself with an intertwiner. Quantisation of $T^*(SU(2))^{d} / SU(2))$ is the full space of $d$-valent intertwiners, $\bigoplus_{j_I} \text{Inv} \otimes_{I = 1}^d \mathcal{H}_{j_I}$. Here Inv$\otimes_{I = 1}^d \mathcal{H}_{j_I}$ is the space of $d$-valent intertwiners with given fixed spins $\{j_I\}$ i.e. given fixed face areas, corresponding to a quantisation of $\mathcal{S}_d$. A collection of neighbouring quantum polyhedra has been associated to a spin network of arbitrary valence \cite{Bianchi:2010gc}, with the labelled nodes and links of the latter being dual to labelled polyhedra and their shared faces respectively.  Then for a quantum tetrahedron, the 1-particle Hilbert space is taken to be
\be
\mathcal{H} = \bigoplus_{j_I} \text{Inv} \otimes_{I = 1}^4 \mcH_{j_I} \;,
\ee
with quantum states of an $N$-particle system belonging to
\be
\mathcal{H}_N = \mathcal{H}^{\otimes N} \;.
\ee
We can equivalently work with the holonomy representation of the same quantum system in terms $SU(2)$ group data, which is also the state space of a single gauge-invariant quantum of a group field theory defined on an $SU(2)^{4}$ base manifold \cite{Oriti:2006se, Oriti:2013aqa},
\be \label{singleH}
\mathcal{H} = L^2(SU(2)^{4}/SU(2)) \;.
\ee
A further, equivalent representation could be given in terms of non-commutative Lie algebra (flux) variables \cite{Baratin:2010wi,Baratin:2010nn}. 

\

As discussed in section \ref{tet}, mechanical models of $N$ quantum tetrahedra can be defined by a set of gluing operators defined on $\mathcal{H}_N$. The general discussion therein is applicable here also. The basic ingredient of gluing is again to define face sharing conditions. For instance, the classical constraints of equation \eqref{gluing} can be implemented by group averaging of wavefunctions \cite{Oriti:2013aqa},  
\be \label{gluing2}
 \Psi_\gamma (\{g_{(nI)}g_{(mJ)}^{-1}\}) = \prod_{(nI,mJ)|_\gamma} \int_{SU(2)} dh_{(nI,mJ)} \; \psi(\{ g_{(nI)}h_{(nI,mJ)} , g_{(mJ)}h_{(nI,mJ)} \}) \ee
where we have used the notation introduced in section \ref{manytet} above, and $\psi \in \mathcal{H}_N$ is a wavefunction for a system of generically disconnected $N$ tetrahedra. So, a wavefunction defined over full links $(nI,mJ)$ of a graph $\gamma$ is a result of averaging over half-links $(nI)$ and $(mJ)$ by $SU(2)$ elements $h_{(nI,mJ)}$. The same can also be implemented in terms of fluxes $X$, using a non-commutative Fourier transform between the holonomy and flux variables \cite{Oriti:2013aqa}. \

Thus a quantum mechanical model of a system of $N$ tetrahedra consists of the unconstrained Hilbert space $\mathcal{H}_N$, an operator algebra defined over it and a set of gluing operators specifying the model.

Now for a quantum multi-particle system, a Fock space is a suitable home for configurations with varying particle numbers. For bosonic\footnote{As for a standard multi-particle system, bosonic statistics corresponds to a symmetry under particle exchange. For the case when a system of quantum tetrahedra is glued appropriately to form a spin network, then this symmetry is interpreted as implementing the graph automorphism of node relabellings.} quanta, each $N$-particle sector is the symmetric projection of the full $N$-particle Hilbert space, so that the Fock space takes the following form,
\be \mathcal{H}_F =  \bigoplus_{N \geq 0} \text{sym}\, \mathcal{H}_N \;.\ee
The Fock vacuum $\ket{0}$ is the one corresponding to a state with no tetrahedron degrees of freedom. \

Then, a system of an arbitrarily large number of quantum tetrahedra is described by the state space $\mathcal{H}_F$, an algebra of operators over it with a special subset of them identified as gluing constraints. Quantum statistical states of tetrahedra are density operators (self-adjoint, positive and trace-class operators) on $\mathcal{H}_F$ \cite{Kotecha:2018gof}. 

\

Let us a consider a system of quantum tetrahedra with a model defined by a (self-adjoint) constraint operator $\h{\mbC}_\gamma$ defined on $\mathcal{H}_F$, and a generalised Gibbs state of the form,
\be\label{qeq} \h{\rho}_{\{\gamma, \beta\}} \propto e^{- \beta \h{\mbC}_\gamma} \ee 
where $\beta$ is the Lagrange multiplier for $\langle \h{\mbC}_\gamma \rangle = 0$. 

%where $\h{\mbG}_\gamma$ is the result of second quantisation of the classical function $G_\gamma$, and $\beta \in \mathbb{R}^{3L}$. This state is the quantum counterpart of \eqref{rhocomp}, where the reduction by half of the temperature dependences is a direct consequence of non-commutation (hence a quantum inter-dependence) between the holonomy and flux variables. Operator $\h{\mbG}_\gamma$ is defined on the full Fock space, but its dependence on a given graph $\gamma$ with a fixed number of vertices $N$, means that it acts on a given $N$-particle subsector of $\mcH$. Its explicit expression in terms of ladder operators is shown in the next subsection.}

%$\h{C}$ defined on $\mathcal{H}_F$, and a generalised Gibbs state of the form,\be\label{qeq} \h{\rho} = \frac{1}{Z} e^{-\beta \h{C}} \;\;,\;\; Z_\beta = \Tr_{\mathcal{H}_F}(e^{-\beta \h{C}}) \ee where $\beta$ is the Lagrange multiplier for $\langle \h{C} \rangle = 0$.

We can further consider contributions from different graphs with a fixed number of vertices, weighted differently with coupling parameters $\lambda$. Such a state takes the form \be \h{\rho}_N = \frac{1}{Z_N(\lambda_\gamma)} e^{-\sum_{\{\gamma\}_N} \frac{1}{\text{Aut}(\gamma)}\lambda_\gamma \h{\mbC}_\gamma} \;. \ee 

Particularly, a density operator with a contribution from a grand-canonical weight of the form $e^{\mu \h{N}}$, corresponds to a statistical state with a varying particle number, where $\h{N}$ is the number operator associated with the Fock vacuum. The corresponding partition function
\be\label{qgrand}
Z(\mu,\lambda_\gamma) = \Tr_{\mcH_F} \left[e^{-\sum_{\{\gamma\}_N} \frac{1}{\text{Aut}(\gamma)}\lambda_\gamma \h{\mbC}_\gamma \;\; + \;\; \mu \h{N} }\right]
\ee
provides the quantum counterpart of the expression \eqref{grand} in \ref{manytet}. As the first term in the exponent is dependent on $N$, we would expect the multiplier $\mu$ to depend on the remaining temperatures, as is also the case in a traditional grand-canonical state. Here since this dependence is non-trivial in general, we leave it separately as above for now. Overall, if an operator $\h{\mbC}$ is the dynamical constraint of the system, which in general could include number- and graph-changing interactions, then one obtains a grand-canonical state of the type above with respect to $\h{\mbC}$.

%The associated equilibrium partition function is then given by
%\be 
%Z_\beta = \Tr_{\mathcal{H}_F}\left[e^{-\beta \left(\sum_a\frac{\beta_a}{\beta}\h{C}_a\right)}\right]. \ee 

%\begin{itemize} 
%\item \textcolor{red}{Example of a simplicial mechanical model of many quantum tetrahedra. Add + discuss coupling constants ---} 
%\end{itemize}

%\textcolor{red}{Consider the following operator as an example,
%\be \label{ex5}
%\h{C}^5 = \int_{SU(2)^{4 \times 2}} \h{\varphi}^*(g^1) K(g^1,g^2) \h{\varphi}(g^2) \;+\; \int_{SU(2)^{4\times 5}} \h{\varphi}(g^1)\h{\varphi}(g^2)\h{\varphi}(g^3)\h{\varphi}(g^4)\h{\varphi}(g^5) V(g^1,g^2,g^3,g^4,g^5) \;+\; c.c.
%\ee
%where $c.c.$ denotes complex conjugate of the second term, and $g^i \equiv (g^i_1,...,g^i_4)$. Kernels $K$ and $V$ may be non-local. For instance, when $V$ has the combinatorics of a 4-simplex like that in \eqref{4simp}, then this $\h{C}_5$ defines a system of quantum tetrahedra which interact to form 4-simplices.} 
%\textcolor{red}{For $V_5$ being simplicial, could write the actual kernels in terms of deltas to make the comparison with the earlier classical example more direct}

%More generally, a quantum statistical description of a collective system of quantum tetrahedra will be given by a partition function associated to an equilibrium quantum statistical state, written with respect to some gluing operator.

%A straightforward example is given by the quantum counterpart of the partition function \eqref{} given in section \ref{}. 
%-------------------------------------------------

\subsection{Field theory of quantum tetrahedra}

The Hilbert space $\mcH_F$\footnote{We remark that $\mathcal{H}_F$ is the GNS representation space of the Fock algebraic state associated with a group field theory Weyl algebra \cite{Kegeles:2017ems, Kotecha:2018gof}.} is generated by a set of ladder operators acting on the cyclic vacuum $\ket{0}$, and satisfying the algebra,
\be \label{ccr} [\h{\varphi}(\vec{g}),\h{\varphi}^*(\vec{g'})] = \delta(\vec{g},\vec{g'})\ee 
where $\delta$ is an identity distribution on the space of smooth, complex-valued $L^2$ functions on $SU(2)^4$, and $\vec{g} \equiv (g_1,...,g_4)$.

This formulation already hints at a second quantised language in terms of quantum fields of tetrahedra. This language can indeed be applied to the whole statistical mechanics framework we have developed, in particular to the partition function obtained in the previous section.

The way to obtain this field-theoretic reformulation is pretty standard. For a state $e^{-\beta \h{C}}$, the traces in the partition function and other observable averages\footnote{We thank Alexander Kegeles for pointing out the relevance of considering the full observable algebra in the present context and helpful discussions.} can be evaluated using an overcomplete basis of coherent states,
\be \ket{\psi} = e^{-\frac{||\psi||^2}{2}}e^{\int d\vec{g}\; \psi(\vec{g}) \h{\varphi}^*(\vec{g})}\ket{0} . \ee 
These states are labelled by $\psi \in \mcH$ and $||.||$ is the $L^2$ norm in $\mathcal{H}$. This gives,
\be \label{coherentZ}
\Tr({e^{-\beta\h{C}} \h{\mcO}}) = \int [D\mu(\psi,\bar{\psi})] \bra{\psi}e^{-\beta\h{C}}\h{\mcO}\ket{\psi} , \;\;\;\;\text{with}\;\; Z = \Tr(e^{-\beta\h{C}} \mathbb{I})
\ee 
where the resolution of identity is $\mbI = \int [D\mu(\psi,\bar{\psi})] \ket{\psi}\bra{\psi}$, and the coherent state functional measure \cite{klauderbook} is,
\be
D\mu(\psi,\bar{\psi}) = \lim_{K \to \infty}\prod_{k=1}^{K}\frac{d\,\text{Re}\psi_k \; d\,\text{Im}\psi_k}{\pi} \;.
\ee
The set of all such observable averages formally defines the complete statistical system. In particular, the quantum statistical partition function can be reinterpreted as the partition function for a field theory (restricted to complex-valued $L^2$ fields) of the underlying quanta, which here are quantum tetrahedra \cite{Oriti:2013aqa}. This can be seen as follows. \

For generic operators $\h{C}(\h{\varphi},\h{\varphi}^*)$ and $\h{\mcO}(\h{\varphi},\h{\varphi}^*)$ as polynomial functions of the generators, and a given (but generic) choice of the operator ordering defining the exponential operator, the integrand of the statistical averages can be treated as follows,
\begin{align}
\bra{\psi}e^{-\beta\h{C}}\h{\mcO}\ket{\psi} &= \bra{\psi}\sum_{k=0}^{\infty}\frac{(-\beta)^k}{k!}  \h{C}^k \h{\mcO}\ket{\psi} \nonumber \\
&= \bra{\psi} :e^{-\beta \h{C}} \h{\mcO}: \ket{\psi} + \bra{\psi} :\text{po}_{{C},{\mcO}}(\h{\varphi},\h{\varphi^*},\beta) : \ket{\psi}
\end{align}
where to get the second equality, we have used the commutation relations \eqref{ccr} on each $\h{C}^k\h{\mcO}$, to collect all normal ordered terms $:\h{C}^k \h{\mcO}:$, %with order 0 in $\hbar$ (and all orders in $\beta$)
to get the normal ordered $:e^{-\beta \h{C}} \h{\mcO}:$, and the second term is a collection of the remaining terms arising as a result of swapping $\h{\varphi}$'s and $\h{\varphi}^*$'s, which will then in general be a normal ordered series in powers of $\h{\varphi}$ and $\h{\varphi^*}$, with coefficient functions of $\beta$. The precise form of this series will depend on both $\h{C}$ and $\h{\mcO}$. %, with minimum order being 1 in $\hbar$. 
Recalling that coherent states are eigenstates of the annihilation operator, $\h{\varphi}(\vec{g})\ket{\psi} = \psi(\vec{g}) \ket{\psi}$, we have
\be
\bra{\psi} :e^{-\beta \h{C}}\h{\mcO}: \ket{\psi} = e^{-\beta C[\bar{\psi},\psi]} \mcO[\bar{\psi},\psi] 
\ee
where $C[\bar{\psi},\psi] = \bra{\psi}\h{C}\ket{\psi}$ and $\mcO[\bar{\psi},\psi] = \bra{\psi}\h{\mcO}\ket{\psi}$. Denoting by operator $\h{A}_{C,\mcO} \equiv \text{po}_{C,\mcO}(\h{\varphi},\h{\varphi^*},\beta)$ we have,
\be
 \bra{\psi} :\h{A}_{C,\mcO}(\h{\varphi},\h{\varphi^*},\beta) : \ket{\psi} = A_{C,\mcO}[\bar{\psi},\psi,\beta]\; ,
\ee
which encodes all higher order quantum corrections\footnote{Notice that $A$ is not necessarily of an exponential form.}. Thus, averages \eqref{coherentZ} can be written as
\be \label{avgs} \Tr({e^{-\beta\h{C}} \h{\mcO}}) = \int [D\mu(\psi,\bar{\psi})] \; \left( e^{-\beta C[\bar{\psi},\psi]} \mcO[\bar{\psi},\psi] + A_{C,\mcO}[\bar{\psi},\psi,\beta] \right) \ee
In particular, the quantum statistical partition function for a dynamical system of complex-valued $L^2$ fields $\psi$ defined on the base manifold $SU(2)^4$ is %an effective field theory..\textcolor{orange}{expand...}.. plus quantum corrections,
\be \label{pf} Z = \int [D\mu(\psi,\bar{\psi})] \; \left(e^{-\beta C[\bar{\psi},\psi]} + A_{C,\mbI}[\bar{\psi},\psi,\beta]\right) \;\equiv\; Z_{0} + Z_{\mcO(\hbar)} 
\ee
where, by notation $\mcO(\hbar)$ we mean only that this sector of the full theory encodes all higher orders in quantum corrections relative to $Z_0$.\footnote{Further investigation into the interpretation, significance and consequences of this rewriting of $Z$ in discrete quantum gravity is left for future work.} 
This full set of observable averages (or correlation functions) \eqref{avgs}, including the above partition function, defines thus a statistical field theory of quantum tetrahedra (or in general, polyhedra with a fixed number of boundary faces), %\textcolor{blue}{with its algebra of observables\footnote{\textcolor{blue}{The resultant functional partition function cannot be understood as defining a model with respect to the original observable algebra. Particularly, the new algebra is not expected to be mapped straightforwardly from the original one via averages in coherent states. That the resultant field system is in general equipped with a different observable algebra is expected, because the original description should in principle have detailed knowledge of the microscopics of the system, while the field theory algebra is constituted by functionals of collective field variables. }} that are generic functionals of the variables $\psi$ and $\bar{\psi}$}.
 characterised by a combinatorially non-local statistical weight, i.e. a group field theory. This derivation and interpretation of the foundation of group field theories was suggested in \cite{Oriti:2013aqa}, and the present work puts it on more solid grounds. If we are able to either reformulate exactly, or under suitable approximations, $A_{C,\mcO}$ in the following way,
 \be A_{C,\mcO} = A_{C,\mbI}[\bar{\psi},\psi,\beta] \,\mcO[\bar{\psi},\psi]  \ee
 then, the partition function \eqref{pf} defines a statistical field theory for the algebra of observables $\mcO[\bar{\psi},\psi]$. Moreover, if we are further able to rewrite $Z$ in terms of a simple exponential measure under some approximations, 
\be
Z_{\text{eff}} = \int [D\mu(\psi,\bar{\psi})] \; e^{- C_{\text{eff}}[\bar{\psi},\psi, \beta, C, A]} 
\ee
then the correspondence with a standard field theory would be even more manifest.  \

The comparison with existing group field theory models, for topological BF theories, thus in absence of additional geometricity conditions and simply using gluing conditions of holonomy-flux data, shows that these are obtained from our statistical construction, but by starting from a reformulation of the initial gluing constraints. Recall that the constraint equation
\be \label{con} \h{\mbC}\ket{\Psi}=0\ee
can be equivalently recast in the form 
\be \label{pro} \h{\mathbb{P}}\ket{\Psi}=\ket{\Psi} \ee
where $\h{\mathbb{P}}\sim \delta(\h{\mbC})$ is the operator projecting on the kernel of $\h{\mbC}$. Physical states are those left invariant by the projector operator (namely $\ket{\Psi} \in Ran(\h{\mathbb{P}})$). In particular, we can recast equation \eqref{pro} in the form of another constraint relation, by considering the complementary operator $\h{\mathbb{Q}} \equiv \mbI -\h{\mathbb{P}}$, hence writing
\be \label{pro2} \h{\mathbb{Q}} \ket{\Psi}=0, \ee
with physical states now being in the kernel of $\h{\mathbb{Q}}$.\footnote{For projectors we have that  $Ran(\h{\mathbb{P}}) = Ker(\mbI -\h{\mathbb{P}} )$ and $Ker(\h{\mathbb{P}}) = Ran(\mbI - \h{\mathbb{P}} )$.} \

To give an example, consider the mechanical model of quantum tetrahedra having the combinatorial structure of the boundary of a dipole as encoded by the classical constraints \eqref{dipcon}. Then, 1st quantising this system results in four operators $\{\h{C}_\ell\}$, each one associated with a different full link of the graph. For instance, for a given full link $\ell$ with classical flux constraint $C_\ell = X_{(nI)} + X_{(mJ)}$, we can define an operator $\h{C}_\ell := \h{X}_{(nI)} \otimes \h{1}_{(mJ)} + \h{1}_{(nI)} \otimes \h{X}_{(mJ)}$, where operators $\h{1}$ and $\h{X}$ are defined on single link space $L^2(SU(2))$. Operator $\h{C}_\ell$ thus acts on two half-link states as follows,
\be \h{C}_\ell \ket{X_{(nI)},X_{(mJ)}} = C_\ell  \ket{X_{(nI)},X_{(mJ)}} \;. \ee
The state of two disconnected tetrahedra (or a dipole that is yet to be made) can be written as,
\be \ket{\psi_2} = \otimes_{\ell=1}^{4} \ket{X_{(nI)},X_{(mJ)}} = \ket{\vec{X}_1,\vec{X}_2} \;,\ee
where $\vec{X}_n \equiv (X_{(n1)},...,X_{(n4)})$, and the second equality is due to bosonic statistics that allows permutations of arguments freely. The 1st quantised operator for the whole graph is, \be \h{C} := \bigotimes_{\ell=1}^{4} \h{C}_\ell \;\;\;\;\;\;,\;\;\;\;\;\; \h{C}\ket{\psi_2} = C \ket{\psi_2} \ee
where $C \equiv (C_1,...,C_{4}) \in \su(2)^{* 4}$. Clearly, $\h{C}$ is a constraint operator which annihilates relevant (here, dipole) states when the classical constraints are satisfied, that is when $C=0$. The corresponding 2nd quantised operator on the Fock space can be defined as,
\begin{align} \h{\mbC} &= \int d\vec{X}_1 d\vec{X}_2 \; \h{\varphi}^*(\vec{X}_1)\h{\varphi}^*(\vec{X}_2) \bra{\vec{X}_1,\vec{X}_2} \h{C} \ket{\vec{X}_1,\vec{X}_2} \h{\varphi}(\vec{X}_1)\h{\varphi}(\vec{X}_2) \nonumber \\
&= \int d\vec{X}_1 d\vec{X}_2 \; \h{\varphi}^*(\vec{X}_1)\h{\varphi}^*(\vec{X}_2)  \; C(\vec{X}_1,\vec{X}_2)  \; \h{\varphi}(\vec{X}_1)\h{\varphi}(\vec{X}_2) \;. \end{align}     
From this, we can define a projector $\h{\mbP} \sim \delta(\h{\mbC})$ as\footnote{The projector operator can equivalently be written in holonomy basis with identical gluing content.}
\begin{align} \label{exP} \h{\mbP} &:= \int d\vec{X}_1 d\vec{X}_2 \; \h{\varphi}^*(\vec{X}_1) \h{\varphi}^*(\vec{X}_2)  \; \prod_{\ell=1}^{4} \delta(C_\ell)  \;\; \h{\varphi}(\vec{X}_1)\h{\varphi}(\vec{X}_2) \nonumber \\
&\equiv \int d\vec{X}_1 d\vec{X}_2 \; \h{\varphi}^*(\vec{X}_1) \h{\varphi}^*(\vec{X}_2)  \; V_{\text{dip}}(\vec{X}_1,\vec{X}_2)  \; \h{\varphi}(\vec{X}_1)\h{\varphi}(\vec{X}_2) \;. \end{align}
Thus the specific choice of classical constraints one wants to impose enter the result in terms of the matrix elements of the corresponding quantum operators in the Fock space. These matrix elements become the convolution kernels $V$ for fields in the projector reformulation. One can then proceed to define a constraint $\h{\mbQ} = \mbI - \h{\mbP}$. \\

Now the example projector \eqref{exP} presented above is number-conserving, evident from equal number of creation and annihilation operators in its expression. It simply picks out those 2-tetrahedra states whose data satisfy the combinatorics of a dipole. Moreover, it is a projector operator on the subsector $\mcH_2$ of $\mcH_F$, in the sense of satisfying $\h{\mbP}^2\ket{\psi} = \h{\mbP}\ket{\psi}$ for all $\psi \in \mcH_2$. General dynamics should include both graph- and number-changing interactions, and the associated projector operator should be a projector on the full $\mcH_F$. Such an operator would thus in general have contributions from all possible $N$-particle subsectors (corresponding to $N$ nodes in the boundary graph) encoding interactions between $m$ `incoming' particles and $n$ `outgoing' particles (with $N = m+n$) \cite{Oriti:2013aqa},
\be \h{\mbP} = \sum_{n,m} \lambda_{n,m} \h{P}_{n,m} \ee
where operators $\h{P}_{n,m}$ (not necessarily projectors on $\mcH_N$) are of the form,
\be \h{P}_{n,m}  = \int d\vec{g}_1...d\vec{g}_n d\vec{h}_1...d\vec{h}_m \; \h{\varphi}^*(\vec{g}_1)...\h{\varphi}^*(\vec{g}_n) V_{n,m}(\vec{g}_1,...,\vec{g}_n,\vec{h}_1,...,\vec{h}_m) \h{\varphi}(\vec{h}_1)...\h{\varphi}(\vec{h}_m) \;. \ee 
Notice that our earlier example of the dipole projector is a special case of the above, specifically $ \h{P}_{2,2} = \h{\mbP} $ (of equation \eqref{exP}), with $V_{2,2}(\vec{g}_1,\vec{g}_2,\vec{h}_1,\vec{h}_2) = \delta(\vec{g}_1,\vec{h}_1)\delta(\vec{g}_2,\vec{h}_2) V_{\text{dip}}(\vec{g}_1,\vec{g}_2)$. \

Other commonly encountered terms are of the form 
\be \h{P}_{0,5} + \h{P}_{5,0} =   \int d\vec{g}_1...d\vec{g}_5 \; V_{\text{4sim}}(\vec{g}_1,...,\vec{g}_5) \; \h{\varphi}(\vec{g}_1)...\h{\varphi}(\vec{g}_5)  +  \h{P}_{0,5}^* \;,\ee
where,
$ V_{\text{4sim}}(\vec{g}_1,...,\vec{g}_5) = \prod_{\ell=1}^{10}\delta(C_\ell) $, with $C_\ell$ being the classical constraints corresponding to 4-simplex combinatorics. In this case then, the approximated statistical field theory as derived from the full quantum statistical system (when the latter is taken to be in a generalised Gibbs state) is
\be
Z_0 = \int [D\mu(\psi,\bar{\psi})] \, e^{-\beta (1 - {P}_{0,5} - \overline{P}_{0,5})} \,,
\ee
with the statistical weight dictated by
\be
({P}_{0,5} + \overline{P}_{0,5})[\bar{\psi},\psi] =  \int_{SU(2)^{4\times 5}} \psi(\vec{g}_1)\psi(\vec{g}_2)\psi(\vec{g}_3)\psi(\vec{g}_4)\psi(\vec{g}_5) V_{\text{4sim}}(\vec{g}_1,...,\vec{g}_5) \;+\; c.c. \;.
\ee
Hence for this choice of gluings and aforementioned approximations to the coherent state averages \eqref{avgs}, the partition function defines a 4d simplicial group field theory model of complex-valued, $SU(2)$-gauge invariant, $L^2$ fields $\psi$, defined on the base manifold $SU(2)^4$. \

For a general projector $\h{\mbP}$ as discussed above with its constraint $\h{\mbQ}$, if we now repeat the derivation of the canonical partition function %\footnote{For a state $e^{-\beta \h{\mbQ}}$, we have $Z = \Tr_{\mcH_F}(e^{-\beta \h{\mbQ}}) = \text{dim}(Ker \h{\mbQ}) + e^{-\beta}\text{dim}(Ran \h{\mbQ}) + \Tr_{\overline{\mathcal{K}}}(e^{-\beta \h{\mbQ}})$, where $\mcH_F = \mathcal{K} \oplus \bar{\mathcal{K}}$ and $\mathcal{K} \equiv Ker \h{\mbQ} \oplus Ran \h{\mbQ}$. Thus, convergence of $Z$ requires convergence of the expression on the right hand side.}
 by considering
\be\label{qgrand}
Z_{\mu, \beta} = \Tr_{\mathcal{H}_F} \left[e^{- \beta\h{\mathbb{Q}} +\mu \h{N} }\right]=\Tr_{\mathcal{H}_F} \left[e^{- \beta(\mbI-\h{\mathbb{P}}) +\mu \h{N} }\right],
\ee
we will end up dealing with a statistical weight expressed in terms of matrix elements of the projector operator, 
\be
\mathbb{P}[\bar{\psi},\psi] = \sum_{n,m} \lambda_{n,m}  \int_{SU(2)^{4(n+m)}} \bar{\psi}(\vec{g}_1) \dots \bar{\psi}(\vec{g}_n) V_{n,m}(\vec{g}_1,\dots, \vec{g}_n, \vec{h}_1,\dots,\vec{h}_m) \psi( \vec{h}_1) \dots \psi( \vec{h}_m) \;.
\ee
This directly gives the group field theory interaction kernels for BF models, which are expressed in terms of products of delta functions whose arguments are the classical gluing constraints.

To summarise, in this statistical formulation we are able to give a more solid foundation to the picture in which a group field theory is a quantum field theory of tetrahedra (or polyhedra, in general), and the kernels of a group field theory action\footnote{What we understand better now as (matrix elements of) gluing constraints and call ${\mbQ}$, is customarily called action, and treated also like a Euclidean action (even though it is not associated with any notion of Wick rotating from a Lorentzian action due to the absence of any spatiotemporal structures of the present system) in group field theory literature.} originate from non-local many-body interactions (gluing constraints) between the underlying quanta.

\section{Conclusion}

We have investigated the statistical mechanics of classical and quantum tetrahedra, which are candidates for quanta of spacetime geometry in discrete quantum gravity approaches. Particularly, we have focused on the definition of Gibbs equilibrium states, in such a background independent context.  They can be defined using Jaynes' principle, which does not rely on the identification of any (time) symmetry or automorphism for characterising the state, but only on the requirement of maximal entropy subject to macroscopic constraints (which are then approximately satisfied in terms of expectation values).  Starting with a system of many classical tetrahedra, we have presented its mechanics and statistical mechanics. As a first illustrative example, we have defined a Gibbs state for the case of the closure constraint for a single classical tetrahedron. Already this example shows that, in a constrained system, a Gibbs or a microcanonical state can be used respectively to partially (on average, $\langle C \rangle = 0$) or exactly ($C=0$) impose the constraints. In other words, the imposition of constraints can be viewed in a novel way in terms of identifying suitable statistical states on the full unconstrained state space. Further, the particular example of a Gibbs state with respect to the closure condition of a tetrahedron is a generalisation of Souriau's Gibbs states to the case of first class constraints. We then consider generalised Gibbs states in a system of (arbitrary) many tetrahedra, with respect to gluing constraints which produce a (approximately, twisted) geometric configuration for connected simplicial complexes, formed by the same tetrahedra. Finally, we have described how our construction translates naturally at the quantum level, in terms of a Hilbert (Fock) space of many quantum tetrahedra and constraint operators acting on them (with the same geometric interpretation). After presenting the quantum statistical mechanics of many tetrahedra, we discuss how the same is recast in the form of a quantum statistical field theory partition function for tetrahedra using a 2nd quantized reformulation and field coherent states (as customary); this corresponds, in fact, to the partition function of a group field theory description of the same system  of many quantum tetrahedra.   \

The statistical framework presented in this work could be used to explore in detail specific examples of simplicial gravity (or group field theory) models with direct or stronger geometric interpretation, and thus of greater interest for quantum gravity. For instance, one could consider the state space of geometric (in the sense of metric) tetrahedra and utilise a generalised Gibbs state to define the partition function with a dynamics encoded by the Regge action. Another interesting direction would be to define a Gibbs density implementing not only gluing constraints but also shape-matching constraints on the twisted geometry space, or simplicity constraints, thus reducing again to a proper Regge geometry from flux-holonomy data. More generally, our framework can be used, starting from any given concrete model, to extract and analyse the thermodynamics and hydrodynamics of the underlying system of (quantum) tetrahedra. It is at this coarse grained level of description, in fact, that we expect a continuum spacetime and geometry, with an approximately gravitational dynamics, to emerge \cite{Oriti:2018tym, Oriti:2018dsg}.   

%-------------------------------------------------

\acknowledgments

IK is grateful to DAAD for funding under the program ``Research Grants - Doctoral Programmes in Germany, 2015-16 (57129429)" during which most of this project was completed. 
\bibliographystyle{unsrt}
\bibliography{refJG}

\end{document}